\documentclass[useAMS,usenatbib]{mn2e}

\usepackage[]{natbib}
\usepackage{graphicx}

\title[Transit Timing of WASP-14\,b]{WASP-14\,b: Transit Timing analysis of 19 light curves\thanks{Based on observations collected at the Centro Astronómico Hispano Alem\'{a}n (CAHA) at Calar Alto, operated jointly by the Max-Planck Institut f\"{u}r Astronomie and the Instituto de Astrofísica de Andaluc\'{i}a (CSIC). Based on observations obtained with telescopes of the University Observatory Jena, which is operated by the Astrophysical Institute of the Friedrich-Schiller-University Jena.}}
\author[St. Raetz et al.]{St. Raetz$^{1,2}$\thanks{E-mail:sraetz@cosmos.esa.int}, G. Maciejewski$^{3}$, M. Seeliger$^{2}$, C. Marka$^{2,4}$, M. Fern\'{a}ndez$^{5}$, \newauthor T. G\"{u}ver$^{6}$, E. G\"{o}\u{g}\"{u}\c{s}$^{7}$, G. Nowak$^{8,9}$, M. Va\v{n}ko$^{10}$, A. Berndt$^{2}$, T. Eisenbeiss$^{2}$, \newauthor M. Mugrauer$^{2}$, L. Trepl$^{2}$, J. Gelszinnis$^{11}$\\
$^{1}$Scientific Support Office, Directorate of Science and Robotic Exploration, European Space Research and Technology Centre (ESA/ESTEC),\\ Keplerlaan 1, NL-2201 AZ Noordwijk, the Netherlands\\
$^{2}$Astrophysikalisches Institut und Universit\"{a}ts-Sternwarte, Schillerg\"{a}\ss{}chen 2-3, D-07745 Jena, Germany\\
$^{3}$Centre for Astronomy, Faculty of Physics, Astronomy and Informatics, Nicolaus Copernicus University, Grudziadzka 5, PL-87-100 Torun,\\ Poland \\
$^{4}$Instituto Radioastronom\'ia Milim\'{e}trica (IRAM), Avenida Divina Pastora 7, E-18012 Granada, Spain \\
$^{5}$Instituto de Astrof\'{\i}sica de Andaluc\'{\i}a, CSIC, Apdo. 3004, E-18080 Granada, Spain\\
$^{6}$Department of Astronomy and Space Sciences, Science Faculty, Istanbul University, Beyaz\i t, 34119 Istanbul, Turkey \\
$^{7}$Faculty of Engineering and Natural Sciences, Sabanc\i \ University, Orhanl\i -Tuzla, 34956 Istanbul, Turkey \\
$^{8}$Instituto de Astrof\'isica de Canarias, C/ v\'ia L\'actea, s/n, E-38205 La Laguna, Tenerife, Spain \\
$^{9}$Departamento de Astrof\'isica, Universidad de La Laguna, Av. Astrof\'isico Francisco S\'anchez, s/n, E-38206 La Laguna, Tenerife, Spain \\
$^{10}$Astronomical Institute, Slovak Academy of Sciences, 059 60 Tatransk\'{a} Lomnica, Slovakia \\
$^{11}$Th\"{u}ringer Landessternwarte Tautenburg, Sternwarte 5, D-07778 Tautenburg, Germany 
             }
\begin{document}

\date{Accepted . Received ; in original form }

\pagerange{\pageref{firstpage}--\pageref{lastpage}} \pubyear{2002}

\maketitle

\label{firstpage}

\begin{abstract}
Although WASP-14\,b is one of the most massive and densest exoplanets on a tight and eccentric orbit, it has never been a target of photometric follow-up monitoring or dedicated observing campaigns. We report on new photometric transit observations of WASP-14\,b  obtained within the framework of `Transit Timing Variations $@$ Young Exoplanet Transit Initiative' (TTV$@$YETI). We collected 19 light-curves of 13 individual transit events using six telescopes located in five observatories distributed in Europe and Asia. From light curve modelling, we determined the planetary, stellar, and geometrical properties of the system and found them in agreement with the values from the discovery paper.\\ A test of the robustness of the transit times revealed that in case of a non-reproducible transit shape the uncertainties may be underestimated even with a wavelet-based error estimation methods. For the timing analysis we included two publicly available transit times from 2007 and 2009. The long observation period of seven years (2007--2013) allowed us to refine the transit ephemeris. We derived an orbital period 1.2\,s longer and 10 times more precise than the one given in the discovery paper. We found no significant periodic signal in the timing-residuals and, hence, no evidence for TTV in the system.
\end{abstract}


\section{Introduction}

The star WASP-14 (\textit{V}\,=\,9.75 mag) was found to have a transiting planet by \citet{2009MNRAS.392.1532J}. The planet orbits the host star with a period of 2.24\,d causing transits with a depth of 11 millimag (mmag) and a duration of 2.8\,h. High-accuracy photometric and spectroscopic follow-up observations allowed one to determine the mass of the planet $M_\mathrm{P}\,=\,7.34\,\pm\,0.50\,M_{\mathrm{Jup}}$, the planet radius $R_\mathrm{P}\,=\,1.28\,\pm\,0.08\,R_{\mathrm{Jup}}$, and the orbital semimajor axis $a\,=\,0.036\,\pm\,0.001$\,au. These findings show that WASP-14\,b is one of the densest exoplanets with an orbital period of less than three days. The stellar density, the effective temperature, the rotation rate, and the high lithium-abundance indicate a very young age between 0.5 and 1.0\,Gyr. All system parameters known from literature are summarized in Table~\ref{Werte_WASP14}.
\begin{table}
\centering
\caption{Physical and orbital properties of the WASP-14\,b system summarized from literature.}
\label{Werte_WASP14}
\begin{tabular}{cr@{\,$\pm$\,}lc}
\hline \hline
Parameter & \multicolumn{2}{c}{Value} & Ref \\ \hline
Epoch zero transit time $T_{0}$  [d] & \multicolumn{2}{c}{2454963.93676} & [1] \\
 & & 0.00025 &  [1] \\
Orbital period $P$  [d] & 2.2437704 & 0.0000028 & [1] \\
Semimajor axis $a$ [au] & 0.036 & 0.001 & [2] \\
Inclination $i$  [$^{\circ}$] & 84.32 & 0.60 & [2] \\
Eccentricity $e$ & 0.087 & 0.002 & [3] \\
Argument of pericentre $\omega$ [$^{\circ}$] & -107.1 & 0.5 & [3] \\
Mass star $M_{\mathrm{A}}$  [M$_{\odot}$] & 1.211 & 0.125 & [2] \\
Radius star $R_{\mathrm{A}}$  [R$_{\odot}$] & 1.306 & 0.070 & [2] \\
Effective temperature $T_{\mathrm{eff}}$  [K] & 6475 & 100 & [2] \\
Surface gravity star log$\,g_{\mathrm{A}}$ & 4.29 & 0.04 & [2] \\
Metallicity $\left[ \frac{M}{H}\right] $ & 0.0 & 0.2 & [2] \\
Mass planet $M_{\mathrm{b}}$  [$M_{\mathrm{Jup}}$] & 7.341 & 0.500 & [2] \\
Radius planet $R_{\mathrm{b}}$  [$R_{\mathrm{Jup}}$] & 1.281 & 0.079 & [2] \\
Distance $d$ [pc] & 160 & 20 & [3] \\ 
Age [Gyr] & \multicolumn{2}{c}{$\sim$\,0.5\,-\,1.0} & [2] \\
Spectral type &  \multicolumn{2}{c}{F5V} & [2] \\
\hline \hline
\end{tabular}
\\ References: [1] \citet{2009PASP..121.1104J}, [2] \citet{2009MNRAS.392.1532J}, and [3] \citet{2013ApJ...779....5B}
\end{table}
\\A very interesting feature of WASP-14\,b reported by \citet{2009MNRAS.392.1532J} and confirmed by \citet{2011MNRAS.413.2500H} is its high orbital eccentricity ($e=0.087\,\pm\,0.002$) for its small orbital distance. At this distance, tidal interactions with the star are expected to circularize the orbit of the transiting planet \citep{1977A&A....57..383Z}. The nearly circular orbits of most close-orbiting exoplanets agree with tidal circularization time-scales significantly shorter than the system age. \citet{2011MNRAS.413.2500H} determined a circularization time-scale for WASP-14\,b to be $5\times10^{7}$ yr, which is significantly shorter than its age. The high eccentricity of WASP-14\,b may thus indicate either a tidal circularization time-scale comparable or longer to the system age or the presence of an additional perturbing body in the system which may significantly delay the process of circularization \citep{2007MNRAS.382.1768M}. It is also possible that the planet may have arrived on its current orbit recently.\\ \citet{2009PASP..121.1104J} have found evidence for a spin--orbit-misalignment ($\lambda\,=\,-33.1^{\circ}\,\pm\,7.4^{\circ}$) in the WASP-14 system by measuring the Rossiter--McLaughlin effect.\\ WASP-14\,b belongs to the class of highly irradiated hot Jupiters. Observations of the thermal emission during three secondary eclipses with \textit{Spitzer} by  \citet{2013ApJ...779....5B} gave first indications about the atmospheric composition and thermal structure of the planet. The observations neither indicate a temperature inversion nor an effective heat exchange between day and night side. The chemical composition is consistent with the solar abundances.\\ WASP-14\,b is a very massive transiting planet. Over 50\% of these massive hot Jupiters show a significant eccentricity while intermediate-mass planets tend to be in circular orbits. Although this finding could be biased because the circularization time-scales with planet mass \citep[][circularization time-scale is longer for massive planets]{2007MNRAS.382.1768M} it may suggest that the formation and evolution of these planets differ from the scenarios for lower mass planets. Therefore, the observation of WASP-14 is particularly interesting to constrain theories of planet formation, migration, and planet--star interaction.\\ Although the non-zero eccentricity and the spin--orbit-misalignment may indicate that there could be additional bodies in the system, WASP-14\,b has never been a target of photometric follow-up observations or observing campaigns, dedicated to detect and characterize signals of transit timing variation (TTV).  \citet{2012MNRAS.426.1291S} re-analysed the data sets of \citet{2009MNRAS.392.1532J} and \citet{2009PASP..121.1104J} and found a significantly different set of physical properties of WASP-14 compared to previous studies. For these reasons, WASP-14\,b was selected as a target of our TTV campaign.  

\section{Observations, data reduction and photometry}

\begin{table*}
\caption{Observatories and instruments which observed transits of WASP-14.}
\label{CCD_Kameras}
\begin{tabular}{cccccccc}
\hline \hline
Observatory & Long. (E) & Lat. (N) & Elevation & Telescope $\diameter$ & Camera & \# Pixel & Pixel scale \\ 
& ($^{\circ}$) & ($^{\circ}$) & (m) & (m) & & & ($''$/pixel) \\ \hline
University Observatory Jena & 11.48 & 50.93 & 370 & 0.25 & CTK$^{a}$ & 1024\,x\,1024 & 2.23 \\
& & & & 0.60 & STK$^{b}$ & 2048\,x\,2048 & 1.55  \\
& & & & 0.25 & CTK-II & 1056\,x\,1027 & 1.19  \\
Calar Alto Observatory & 357.45 & 37.22 & 2168 & 2.20 & CAFOS$^{c}$ & 2048\,x\,2048 & 0.47 \\
Star\'{a} Lesn\'{a} Observatory & 20.29 & 49.15 & 785 & 0.50 & ST10XE & 2184\,x\,1472 & 0.56  \\
Observatorio de Sierra Nevada & 356.62 & 30.06 & 2896 & 1.50 & VersArray: &  &   \\
(OSN) & & & & & 2048B & 2048\,x\,2048 & 0.23  \\
T\"{U}BITAK National Observatory & 30.34 & 36.82 & 2485 & 1.00 & SI 1100 Cryo & 4096\,x\,4097 & 0.31 \\
\hline \hline
\end{tabular}
\\
$^{a}$\citet{2009AN....330..419M}, $^{b}$\citet{2010AN....331..449M}, $^{c}$\citet{1994S&W....33..516M}
\end{table*}

\begin{table*}
\caption{Summary of the WASP-14\,b observations in the period from 2009 April to 2013 April: $N_{\mathrm{exp}}$ -- number of exposures, $T_{\mathrm{exp}}$ -- exposure times, $\Gamma$ -- median number of exposures per minute, pnr -- photometric noise rate.}
\label{Beobachtungslog_WASP14}
\begin{tabular}{lcccccccc}
\hline \hline
Date & Epoch$^{a}$ & Observatory$^{b}$ & Filter & $N_{\mathrm{exp}}$ & $T_{\mathrm{exp}}$ (s) & $rms$ (mmag) & $\Gamma$ & pnr  \\\hline
2009 Apr. 01 & 205 & Jena-CTK & \textit{I} &324 & 60, 50, 40 & 5.16 & 0.87 & 5.54 \\
2009 Apr. 19 & 213 & Jena-CTK &  \textit{I} &320 & 30 & 5.16 & 1.02 & 5.11 \\
2009 May 07 & 221 & Jena-CTK &  \textit{R} & 320 & 25, 20 & 10.03 & 1.22 & 9.08 \\
2011 Feb. 12 & 509 & OSN &  \textit{R} & 1215 & 10 & 3.56 & 0.71$^{c}$ & 4.23 \\
2011 Mar. 02 & 517 & Jena-STK* &  \textit{R} & 395 & 30 & 3.27 & 1.39 & 2.77 \\
& & Jena-CTK-II* &  \textit{R} & 366 & 40 & 5.00 & 1.39 & 4.24 \\
& & Calar Alto* &  \textit{B} &144 & 30 & 6.07 & 1.09 & 5.83 \\
2011 Mar. 11  & 521 & Jena-STK &  \textit{R} & 330 & 45 & 2.46 & 1.10 & 2.34 \\
2011 Mar. 20 & 525 & Jena-STK &  \textit{R} & 360 & 30 & 1.87 & 1.39 & 1.59 \\
& & Jena-CTK-II &  \textit{V} & 535 & 30, 25 & 4.65 & 2.17 & 3.16 \\
& & Calar Alto &  \textit{B} &322 & 30 & 10.36 & 1.10 & 9.87 \\
2011 Mar. 29  & 529 & Jena-STK &  \textit{R} & 401 & 30 & 1.25 & 1.39 & 1.06 \\
& & Star\'{a} Lesn\'{a} &  \textit{R} & 696 & 30 & 3.19 & 3.02 & 1.84 \\
2011 Apr. 07 & 533 & OSN &  \textit{R} & 570 & 30 & 4.10 & 0.34$^{c}$ & 7.08 \\
2011 Apr. 16 & 537 & Calar Alto &  \textit{V} & 404 & 25 & 9.99 & 1.20 & 9.13 \\
2012 Feb. 06 & 669 & Jena-STK* &  \textit{R} & 310 & 30 & 4.21 & 1.39 & 3.57 \\
& & Jena-CTK-II* &  \textit{V} & 208 & 60 & 6.58 & 0.95 & 6.75 \\
2012 Feb. 24  & 677 & Calar Alto &  \textit{V} & 350 & 30 & 0.90 & 0.39$^{c}$ & 1.44 \\ 
2013 Apr. 30  & 869 & T\"{U}BITAK & \textit{R} & 294 & 30 & 3.34 & 0.75 & 3.85 \\
\hline \hline
\end{tabular}
\\
$^{\ast}$partial transit or gaps in the LCs. \\
$^{a}$calculated using the ephemeris in \citet{2009MNRAS.392.1532J}.\\
$^{b}$for a description see Table~\ref{CCD_Kameras}. For the University Observatory Jena also the used camera is listed to avoid confusion. \\
$^{c}$for the final binned LC.
\end{table*}

First observations of WASP-14 were already carried out in 2009 at the University Observatory Jena. Since 2011 February WASP-14\,b has been a target of our TTV campaign. Altogether we collected 19 light curves (LCs) of 13 individual transit events. We used six telescopes (with apertures ranging from 0.25\,--\,2.2\,m) located in five observatories distributed in Europe and Asia. The participating observatories with their telescopes and instruments are summarized in Table~\ref{CCD_Kameras}.\\ The photometric data were reduced by the standard procedures including subtraction of bias and dark and dividing by a sky flat-field. We calibrated the CCD images using the \begin{scriptsize}IRAF\end{scriptsize}\footnote{\begin{scriptsize}IRAF\end{scriptsize} is distributed by the National Optical Astronomy Observatories, which are operated by the Association of Universities for Research in Astronomy, Inc., under cooperative agreement with the National Science Foundation.} routines \textit{darkcombine}, \textit{flatcombine}, and \textit{ccdproc}.\\ Aperture photometry was performed with a modified version of the standard \begin{scriptsize}IRAF\end{scriptsize} routine \textit{phot}. Differential magnitudes were calculated using the method of the optimized artificial comparison star developed by \citet{2005AN....326..134B}. All available field stars are combined to create an optimiszd artificial comparison star by taking a weighted average. Very faint or variable stars get a low weight while bright and constant stars dominate the artificial comparison star.  The final LC is produced by the comparison of WASP-14 with this artificial standard star. \\ Ten different aperture radii were tested. The aperture that produced LCs with the smallest scatter (smallest root mean square, rms) was finally chosen. \\ As a preparation for the LC analysis we fitted the LCs with the \begin{scriptsize}JKTEBOP\end{scriptsize} code \citep{2004MNRAS.349..547S,2004MNRAS.351.1277S}, which is based on the \begin{scriptsize}EBOP\end{scriptsize} program \citep{1981psbs.conf..111E,1981AJ.....86..102P}. It allows us to remove photometric trends (caused by a colour difference between target and comparison star, differential extinction, changes in airmass during the observations) by fitting polynomials up to fifth order simultaneously to the modelling. Throughout this work the LCs were detrended by fitting a second-order polynomial. \\ Finally the differential magnitudes were transformed into fluxes and divided by the average out-of-transit value in order to normalize the LCs. All 19 LCs of WASP-14 are shown at the end of the paper. To give an estimate of the varying quality of the observed LCs, we computed the photometric noise rate \citep[pnr;][]{2011AJ....142...84F}. The pnr is calculated by dividing the rms of each LC, which is a result of the LC fitting with \begin{scriptsize}JKTEBOP\end{scriptsize}, by the square root of the median number of exposures per minute ($\Gamma$), 
\begin{equation}
\mathrm{pnr}=\frac{\mathrm{rms}}{\sqrt{\Gamma}}
\end{equation}
A summary of all observations is given in Table~\ref{Beobachtungslog_WASP14}.

\subsection{University Observatory Jena photometry}

Most observations were carried out at the University Observatory Jena which is located close to the village Gro{\ss}schwabhausen, 10\,km west of the city of Jena. There we have three telescopes (90, 25, and 20\,cm) on the same mount each equipped with an optical CCD camera. For our transit observations we used the 90\,cm Schmidt telescope (60\,cm free aperture in Schmidt-mode) with the CCD-camera STK \citep{2010AN....331..449M} and the 25\,cm Cassegrain auxiliary telescope with the CCD-camera CTK \citep{2009AN....330..419M}. In 2010 August, the CTK was replaced with a new CCD camera called CTK-II (\textit{Cassegrain Teleskop Kamera-II}). The properties of the camera are given in Table~\ref{CTKII}. \\ Between 2009 March and 2012 March, we observed 11 LCs of eight individual transits of WASP-14 at the University Observatory Jena. Six of these eight transit are fully covered. The remaining two transits show gaps in the data due to passing clouds. For the transit from 2012 February 6 only a part of the flat bottom is missing, so that the shape of the transit is still identifiable. The situation is different for the transit from 2011 March 2. In this case the whole egress was lost due to clouds that makes it difficult to determine a precise transit time. \\ Observations were performed in Johnson $V$, Cousins $R$ or $I$ filters. The optics were defocused slightly for the observations with the STK. The exposure times varied from 20 to 60\,s with the 25\,cm Cassegrain telescope and from 30 to 45\,s with the 90/60\,cm Schmidt telescope, depending on weather conditions, airmass, and telescope defocusing.

\begin{table}
\caption{Camera facts of the CTK-II}
\label{CTKII}
\begin{tabular}{lr}\hline \hline
Parameter & Value \\  \hline
Detector: & E2V CCD47-10 \\
Pixel: & 1056\,$\times$\,1027 \\
Pixelscale: & (1.1956 $\pm$  0.0001)\,arcsec\,pixel$^{-1}$ \\
Field of view: & 21.0\,arcmin\,$\times$\,20.4\,arcmin \\
Filter: & Bessell $B, V, R, I$, Gunn $z$ \\
Focus: & Cassegrain \\
\hline \hline
\end{tabular}
\end{table}

\subsection{TTV$@$YETI photometry}

The observation of a large number of individual transit of a planet in front of its host star by a single observatory is difficult. Due to the weather conditions in Central Europe and the orbital period of the transiting planet, it is almost impossible to observe consecutive transits.\\ In 2009, we launched an international observation campaign which is dedicated to the detection and characterization of TTVs for carefully selected transiting planets. The programme is realized by the observation with globally distributed telescopes at different longitudes and is based on the cooperation in the framework of the `Young Exoplanet Transit Initiative' \citep[YETI,][]{2011AN....332..547N}. A description of `TTV$@$ YETI' is available at the website of the project\footnote{http://www.home.umk.pl/~gmac/TTV}.

\subsubsection{Calar Alto 2.2\,m telescope}

Between 2011 February and April, we awarded 2.5 nights (5\,x\,0.5 nights, project F11-2.2-007) with the Calar Alto Faint Object Spectrograph (CAFOS) at the Calar Alto Observatory. Due to extremely bad weather in winter 2011 two out of five transits were lost completely and one was only observed partially. The remaining two events could be observed but also under bad conditions like thin clouds, fog, and full moon. Therefore, the data have a relatively low quality which is not sufficient to measure precise transit times. One additional LC was observed as back-up of project F12-2.2-009 on 2012 February 24.\\ For the observations we used CAFOS in imaging mode and 2\,$\times$\,2 binning. We windowed the field of view to $7.9$\,arcmin\,$\times5.2$\,arcmin to shorten the read-out time. The observed field included WASP-14 and three suitable comparison stars. Because WASP-14 is a relatively bright star the observations were carried out in Johnson $B$ or $V$ band. It turned out that the best LCs could be observed in the $V$ band. To minimize random and flat-fielding errors, we defocused the telescope significantly. The telescope was autoguided during the observations. Depending on the atmosphere transparency and telescope defocusing, we used exposure times of 25 or 30\,s. For the observations from 2012 February 24, we combined three measurements into one data point to obtain a binned light curve.

\subsubsection{Observatorio de Sierra Nevada}

Two complete transit LCs (2011 February 12 and 2011 April 7) were obtained at the Observatorio de Sierra Nevada using the 1.5\,m reflector. With a VersArray:2048B CCD camera with 2048\,$\times$\,2048 pixels and a pixel scale of 0.23\,arcsec\,pixel$^{-1}$, we could observe a field of view of $7.85$\,arcmin\,$\times$\,$7.85$\,arcmin. The exposure times of the $R$-band observations were chosen between 10 and 30\,s. While the first LC were obtained under good weather conditions, the second transit suffered from passing clouds in the first half of the observation. To reduce the scatter, we binned the LCs by averaging every five data points.

\subsubsection{T\"{U}BITAK National Observatory}

One transit that occurred in 2013 April 30 was observed using the Spectral Instruments SI1100 series 4096$\times$4096 CCD camera mounted on the 1.0 m Telescope (T100) at T\"{U}BITAK National Observatory (TUG) in Turkey. The images were acquired in $R$ band with an exposure of 15\,--\,30\,s under good weather conditions.

\subsubsection{Star\'{a} Lesn\'{a} Observatory}

One additional LC of WASP-14 was observed at the Star\'{a} Lesn\'{a} Observatory in Slovakia. The observation in the night of 2011 March 29 was performed with a 50\,cm Newtonian telescope equipped with an SBIG ST10 MXE CCD camera. The CCD-chip consists of 2148\,$\times$\,1472 6.8 $\mathrm{\mu}$m pixels and has a pixel scale of 0.56\,arcsec\,pixel$^{-1}$ corresponding to the field of view of about 24$\times$16 arcmin.The LC was obtained in Cousins $R$ band with an exposure time of 20\,s. \\ Unlike all other observations used in this study, the data reduction and photometry for this transit was already carried out at the Star\'{a} Lesn\'{a} Observatory. The standard correction procedure (bias, dark and flat-field correction) and subsequently aperture photometry was performed by \begin{scriptsize}C-MUNIPACK\end{scriptsize} package\footnote{http://c-munipack.sourceforge.net/}. Since the \begin{scriptsize}C-MUNIPACK \end{scriptsize} package is also based on the \begin{scriptsize}DAOPHOT\end{scriptsize} program \citep{1987PASP...99..191S} like \begin{scriptsize}IRAF\end{scriptsize}, the results of both routines are comparable. To generate an artificial comparison star, at least 20\,--\,30\,\%  of stars with the lowest LC scatter were selected iteratively from the field stars brighter than 2.5\,--\,3 mag below the saturation level. To measure instrumental magnitudes, various aperture radii were used. The aperture which was found to produce LC with the smallest scatter was applied for generation of final LC. Due to problems with the flat-field and because the observations were acquired without autoguiding the LC contains correlated noise.

\section{Light-curve analysis}
\label{lc_analysis}

\begin{table}
\centering
\caption{System parameters resulting from the simultaneous wavelet-based red noise MCMC analysis of the four best quality LCs.}
\label{tbl:TAPmcmc1}
\renewcommand{\arraystretch}{1.1}
\begin{tabular}{lc}
\hline \hline
Parameter & Value \\ \hline
         Period & 2.2437704*\\
    Inclination & 85.3 $^{+1.8}_{-1.1}$\\
           $a$/$R_{\mathrm{A}}$ & 5.98 $^{+0.42}_{-0.32}$\\
          $R_{\mathrm{b}}$/$R_{\mathrm{A}}$ & 0.0965 $^{+0.0021}_{-0.0027}$\\
      Linear LD (\textit{R} band) & 0.39 $^{+0.29}_{-0.24}$ \\
        Quad LD (\textit{R} band) & 0.18 $^{+0.38}_{-0.45}$ \\      
      Linear LD (\textit{V} band) & 0.53 $^{+0.28}_{-0.30}$\\
        Quad LD (\textit{V} band) & 0.12 $^{+0.41}_{-0.40}$\\
           Eccentricity & 0.087*\\ \hline \hline
\end{tabular}
\\
$^{\ast}$Value fixed in MCMC analysis.
\end{table}

\begin{figure}
  \centering
  \includegraphics[width=0.33\textwidth,angle=270]{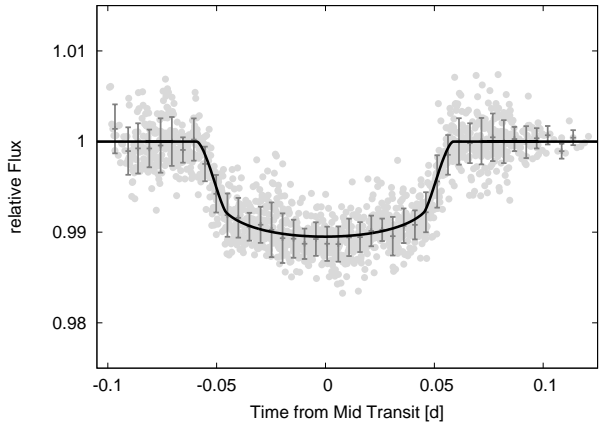}
  \caption{Phase-folded LC of the four best LCs with a pnr\,$<$\,2 mmag that were used to create the template shown as solid black line. The dark grey points show the same fluxes binned in phase, with a bin size of 0.005\,d.}
  \label{alltransit_phased_lc}
\end{figure}

To refine the parameters of the system and to determine precise mid-transit times, it is necessary to model the individual LCs. Therefore, we used the Transit Analysis Package\footnote{http://ifa.hawaii.edu/users/zgazak/IfA/\begin{scriptsize}TAP\end{scriptsize}.html} \citep[\begin{scriptsize}TAP\end{scriptsize} v2.1;][]{2012AdAst2012E..30G}, which employs Markov Chain Monte Carlo (MCMC) techniques to fit transit LCs using the \citet{2002ApJ...580L.171M} model. To calculate the model, \begin{scriptsize}TAP\end{scriptsize} uses the fast exoplanetary fitting code \begin{scriptsize}EXOFAST\end{scriptsize} which was developed by \citet{2013PASP..125...83E}. \begin{scriptsize}TAP\end{scriptsize} has some major advantages. First, it incorporates the wavelet-based technique of \citet{2009ApJ...704...51C} which allows one to estimate more robust parameter uncertainties than classic $\chi^{2}$ methods because it parameterizes uncorrelated as well as time-correlated noise. Secondly, it is able to simultaneously analyse multiple transits observed in different conditions (instrument, filter, weather, etc). Finally, the \begin{scriptsize}TAP\end{scriptsize} code employs the quadratic limb darkening (LD) law which is a better choice to represent the observations than a simple linear law as shown, for example, in \citet{2014MNRAS.444.1351R}. The theoretical LD coefficients that were used as initial values for the fitting were bilinearly interpolated (in effective temperature and surface gravity) from the tables by \citet{2000A&A...363.1081C} using the stellar parameters in Table~\ref{Werte_WASP14}. \\ The LC analysis 
was carried out as explained in \citet[][a]{2013A&A...551A.108M} and \citet[][b]{2013AJ....146..147M}. We selected the best-quality LCs sorted by their pnr for a simultaneous fit using 10 MCMC chains with $10^{5}$ steps each. The orbital inclination $i$ , the semimajor-axis scaled by stellar radius $\frac{a}{R_{\mathrm{A}}}$, and the planetary to stellar radii ratio $\frac{R_{\mathrm{b}}}{R_{\mathrm{A}}}$ were linked together for all LCs. We also connected the LD coefficients $u$ and $v$ but only for the LCs that were observed in the same filter. The orbital period $P$, the eccentricity $e$ and the argument of periastron $\omega$ were kept fixed while the mid-transit times $T_{\mathrm{c}}$, the airmass slopes, and the flux offsets were allowed to vary separately. The selection of the best LCs was done iteratively. The fitting procedure started with a few LCs with small pnr and was repeated several times after adding more LCs with higher pnr. The final selection consists of the four best LCs with a pnr\,$<$\,2 mmag, three LCs in \textit{R} and one in the \textit{V} band. Including transits with a higher pnr degraded the quality of the fit. A phase-folded LC of these four transits including the best-fitting model that correspond to the resulting system parameters given in Table~\ref{tbl:TAPmcmc1} are shown in Fig.~\ref{alltransit_phased_lc}. 

\section{Physical properties}

\begin{figure}
  \centering
  \includegraphics[width=0.33\textwidth,angle=270]{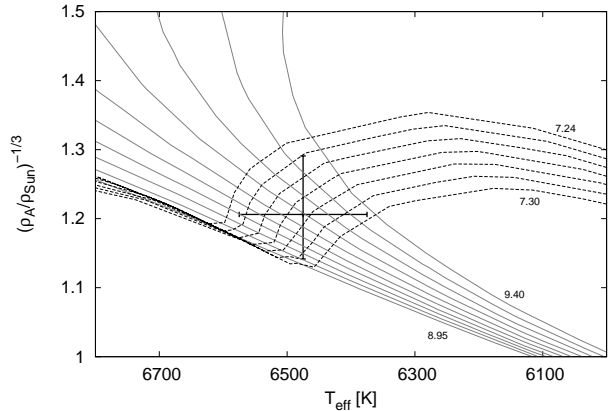}
  \caption{Position of WASP-14 in the $\rho_{\mathrm{A}}^{-1/3}\,-\,T_{\mathrm{eff}}$ plane. The PARSEC isochrones of solar metallicity for log(age)\,=\,7.24\,-\,7.30 with steps of 0.01 and log(age)\,=\,8.95\,-\,9.40 with steps of 0.05 for the very young age and the young age, respectively, are also shown.}
  \label{HRD}
\end{figure}

The results given in Table~\ref{tbl:TAPmcmc1} allow us to calculate stellar, planetary, and geometrical parameters. The mean stellar density $\rho_{\mathrm{A}}$ can be derived directly from the parameters obtained from the light-curve modelling using
\begin{equation}
\label{density}
\rho_{\mathrm{A}}=\frac{3\mathrm{\pi}}{GP^{2}}\left( \frac{a}{R_{\mathrm{A}}}\right)^{3}
\end{equation}
\citep{2010exop.book...55W}, where $G$ is the gravitational constant. To determine the stellar mass we used the $T_{\mathrm{eff}}$ from Table~\ref{Werte_WASP14} and the stellar density to plot WASP-14 into a modified version of the Hertzsprung--Russel diagram (HRD), together with PARSEC isochrones \citep[version 1.2S;][]{2012MNRAS.427..127B}. The result for $\left[ \frac{M}{H}\right]=0.0$ is shown in Fig.~\ref{HRD}. WASP-14 is in an area of the HRD with overlapping isochrones of very young ($\sim$\,20\,Myr) and young ($\sim$\,1\,Gyr) ages. Since \citet{2009MNRAS.392.1532J} estimated an age range of  0.5\,-\,1.0\,Gyr based on lithium abundance, the rotation velocity and a comparison with the models of \citet{2007ApJ...659.1661F}, we excluded the very young ages. Taken also the uncertainty in the metallicity into account the stellar evolutionary models yielded a stellar mass of $M_{\mathrm{A}}=1.30\pm0.06\,\mathrm{M_{\odot}}$ which is consistent with the mass given in \citet{2009MNRAS.392.1532J}. The improved period (see Section \ref{Transit_times}), the stellar mass, the orbital inclination $i$, the eccentricity $e$, and the amplitude of the star's radial velocity $K_{\mathrm{A}}$ taken from \citet{2013ApJ...779....5B} allowed us to redetermine the planetary mass $M_{\mathrm{b}}$. To calculate the semimajor axis $a$, we inserted the orbital period and the masses of star and planet into Kepler's third law. By resolving $a$/$R_{\mathrm{A}}$ and $R_{\mathrm{b}}$/$R_{\mathrm{A}}$ using the already determined value for $a$ we deduce values for $R_{\mathrm{A}}$ and $R_{\mathrm{b}}$. The planetary radius as well as the mass were used to calculate the planetary density $\rho_{\mathrm{b}}$. The impact parameter $b$ which depends on the eccentricity $e$ were calculated using 
\begin{equation}
\label{Impakt_exz}
b=\frac{a\,\mathrm{cos}\,i}{R_{\mathrm{A}}}\frac{1-e^{2}}{1+e\,\mathrm{cos}\,\omega}.
\end{equation}
The surface gravities of star and planet $g_{\mathrm{A}}$ and $g_{\mathrm{b}}$ were calculated using the formulae
\begin{equation}
\label{gb}
g_{\mathrm{b}}=\left( \frac{2\mathrm{\pi}}{P}\right) \frac{(1-e^{2})^{1/2}}{\mathrm{sin}\,i}\frac{a^{2}K_{\mathrm{A}}}{R_{\mathrm{b}}^{2}}
\end{equation}
and
\begin{equation}
\label{gA}
g_{\mathrm{A}}=\left( \frac{2\mathrm{\pi}}{P}\right) \frac{(1-e^{2})^{1/2}}{\mathrm{sin}\,i}\frac{a^{2}K_{\mathrm{b}}}{R_{\mathrm{A}}^{2}}
\end{equation}
with
\begin{equation}
\label{Kb}
K_{\mathrm{b}}=\frac{2\mathrm{\pi} a M_{\mathrm{A}}\mathrm{sin}\,i}{(M_{\mathrm{A}}+M_{\mathrm{b}})P\sqrt{1-e^{2}}}
\end{equation}
\citep{2009MNRAS.394..272S}, where $K_{\mathrm{A}}$ and $K_{\mathrm{b}}$ are the  amplitude of the star's and planet's radial velocity, respectively.
In addition, we calculated the equilibrium temperature of the planet $T_{\mathrm{eq}}$ \citep[assuming a Bond albedo\,=\,0 and only little energy redistribution across the surface of the planet;][]{2007ApJ...671..861H} and the Safronov number $\Theta$ \citep{1972epcf.book.....S}, the square of the ratio of escape velocity of the planet $v_{\mathrm{esc}}$ and orbital velocity $v_{\mathrm{orb}}$:
\begin{equation}
\label{Safronov}
\Theta=\frac{1}{2}\left(\frac{v_{\mathrm{esc}}}{v_{\mathrm{orb}}}\right)^{2}=\frac{a}{R_{\mathrm{b}}}\frac{M_{\mathrm{b}}}{M_{\mathrm{A}}}
\end{equation}
The results of the calculations are summarized in Table~\ref{phys_prop_WASP14}. Our results are consistent with the values in \citet{2009MNRAS.392.1532J} but have larger error bars. This is not surprising taking into account the quality of our ground-based LCs. The physical properties given in \citet{2012MNRAS.426.1291S}, however, are significantly different from our findings. The reason for this discrepancy lies in the interpretation of the results of the LC analysis. \citet{2012MNRAS.426.1291S} modelled two available LCs and obtained a set of solutions (for different LD laws and for fitting either one or two LD coefficients) which was averaged to a final solution. Since both LCs yielded different results, the final derived averaged properties deviate from the already published values. 

\begin{table}
\centering
\caption{Physical properties of the WASP-14 system derived from LC modelling. Values derived by \citet[][J09]{2009MNRAS.392.1532J} and \citet[][S12]{2012MNRAS.426.1291S} are given for comparison.}
\label{phys_prop_WASP14}
\renewcommand{\arraystretch}{1.1}
\begin{tabular}{lr@{\,$\pm$\,}lr@{\,$\pm$\,}lr@{\,$\pm$\,}l}
\hline \hline
 Parameter & \multicolumn{2}{c}{This work} & \multicolumn{2}{c}{J09} & \multicolumn{2}{c}{S12} \\ \hline \hline
& \multicolumn{6}{c}{Planetary parameters} \\ \hline 
$R_{\mathrm{b}}$  [R$_{\mathrm{Jup}}$] & 1.240 & $^{0.116}_{0.103}$ & 1.281 & $^{0.075}_{0.082}$ & 1.633 & 0.092 \\
$M_{\mathrm{b}}$  [M$_{\mathrm{Jup}}$] & 7.59 & $^{0.24}_{0.23}$ & 7.34 & 0.50 & 7.90 & 0.46 \\
$\rho_{\mathrm{b}}$  [$\mathrm{\rho}_{\mathrm{Jup}}$] & 3.73 & $^{1.05}_{0.93}$ & 3.50 & $^{0.64}_{0.50}$ & 1.69 & 0.25 \\
log\,$g_{\mathrm{b}}$ & 4.090 & $^{0.080}_{0.071}$ & 4.010 & $^{0.049}_{0.042}$ & 3.866 & 0.042 \\
$T_{\mathrm{eq}}$ [K] & 1872 & $^{29}_{29}$ & 1866 & $^{37}_{42}$ & 2090 & 59 \\ 
$\Theta$ & 0.345 & $^{0.037}_{0.037}$ & \multicolumn{2}{c}{} & 0.265 & 0.015 \\ \hline
& \multicolumn{6}{c}{Stellar parameters} \\ \hline 
$R_{\mathrm{A}}$  [R$_{\mathrm{\odot}}$] & 1.318 & $^{0.095}_{0.073}$ & 1.306 & $^{0.066}_{0.073}$ & 1.666 & 0.097 \\
$M_{\mathrm{A}}$  [M$_{\mathrm{\odot}}$] & 1.300 & 0.060 & 1.211 & $^{0.127}_{0.122}$ & 1.350 & 0.120 \\
$\rho_{\mathrm{A}}$  [$\mathrm{\rho}_{\mathrm{\odot}}$] & 0.570 & $^{0.120}_{0.092}$ & 0.542 & $^{0.079}_{0.060}$ & 0.293 & 0.042 \\ 
log\,$g_{\mathrm{A}}$ & 4.312 & $^{0.061}_{0.047}$ & 4.287 & $^{0.043}_{0.038}$ & 4.126 &  0.042 \\
$L_{\mathrm{A}}$  [L$_{\mathrm{\odot}}$] & 0.435 & 0.085 & \multicolumn{2}{c}{} & \multicolumn{2}{c}{} \\\hline
& \multicolumn{6}{c}{Geometrical parameters} \\ \hline 
$a$  [au] & 0.037 & 0.001 & 0.036 & 0.001 & 0.037 & 0.001 \\
$i$  [$^{\circ}$] & 85.30 & $^{+1.80}_{-1.10}$ & 84.32 & $^{0.67}_{0.57}$ & 81.1 & 1.5 \\
$b$ & 0.499 & $^{0.194}_{0.120}$ & 0.535 & $^{0.031}_{0.041}$ & 0.752 & 0.133* \\ \hline \hline
\end{tabular}
\\
$^{\ast}$This parameter is not given in \citet{2012MNRAS.426.1291S} but was calculated from $R_{\mathrm{A}}$, $a$, $i$
\end{table}

\section{Transit Timing}
\label{Transit_times}

The mid-transit times were determined by applying the transit model created in Section~\ref{lc_analysis} to the individual LCs using \begin{scriptsize}TAP\end{scriptsize}. The initial parameters for the fitting were the ones given in Table~\ref{tbl:TAPmcmc1}. The mid-transit time, as well as the flux slope and intercept were always a free parameter while the orbital period $P$ and the eccentricity $e$ were kept fixed. The best-model parameters and their uncertainties were set as Gaussian priors and $i$, $a$/$R_{\mathrm{A}}$, $R_{\mathrm{b}}$/$R_{\mathrm{A}}$, and the LD coefficients of each transit were allowed to vary within this range. With this approach we assured that the best-model uncertainties are included in the error bars of the mid-transit time. Several LCs were observed in a filter that has not contributed to the template LC. In those cases the theoretical values $\pm0.5$ was used as Gaussian prior for the LD coefficients. Ten chains of a length of $10^{5}$ steps were used for the MCMC analysis of each LC. Four transits were observed with more than one telescope. These LCs were fitted simultaneously to increase the timing precision. The times have been converted into the barycentric Julian Date based on the barycentric dynamic time (BJD$_{\mathrm{TDB}}$) using the online converter\footnote{http://astroutils.astronomy.ohio-state.edu/time/utc2bjd.html} by \citet{2010PASP..122..935E}. 

\begin{figure}
\begin{minipage}[]{0.45\textwidth}
  \centering
  \includegraphics[height=0.32\textheight, angle=270]{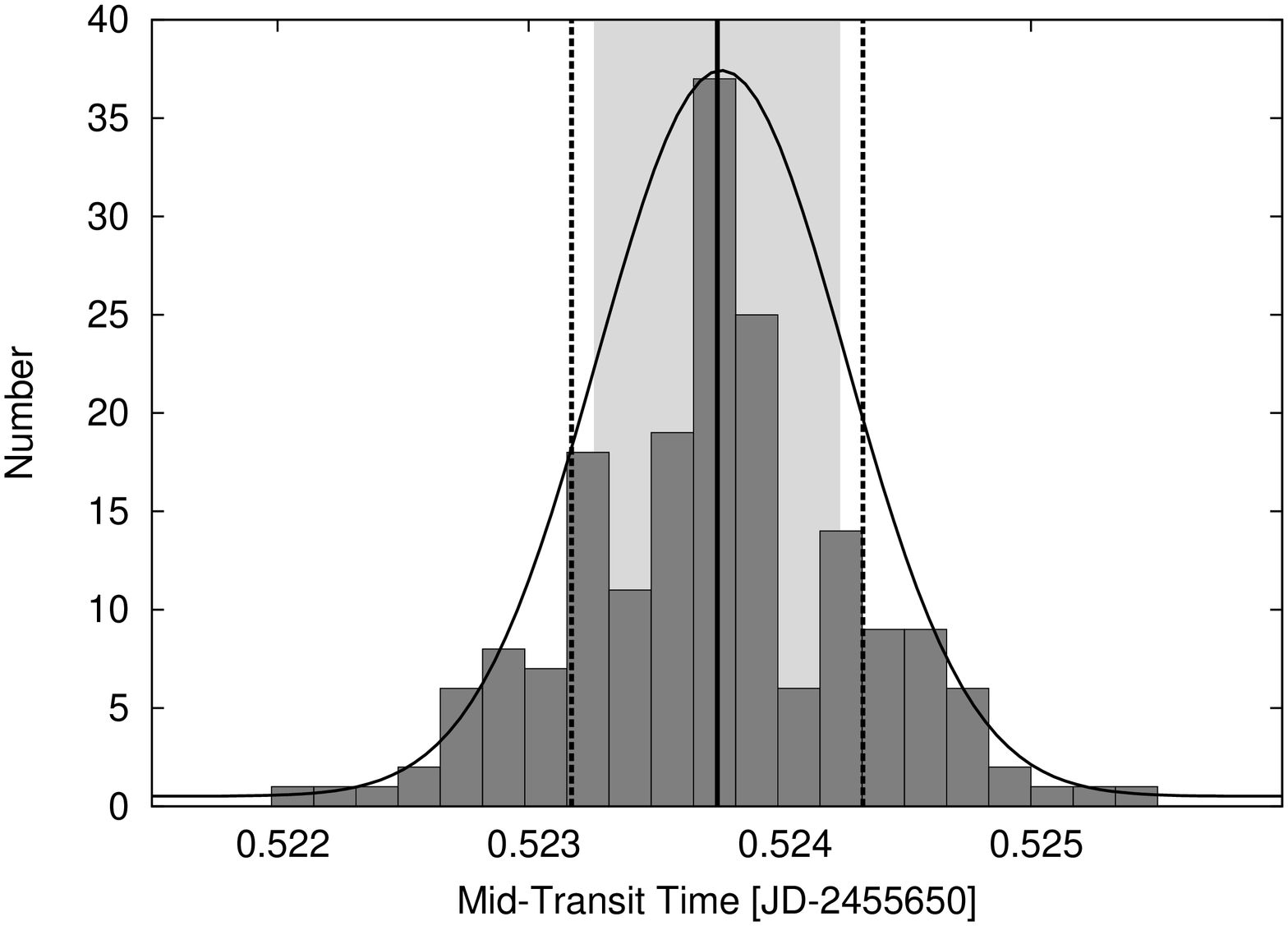}
  \caption{Distribution of the transit times obtained from the analysis of 185 LCs that were created as a transit time robustness check (see text for details) for the transit at epoch 529 (2011 March 29). The thick solid line gives the result of TAP only for the initial LC, the dashed lines give the TAP error bars. The width of the transit time distribution, shown here as grey shaded area, is nicely reproduced by the TAP uncertainties (ratio between distribution width and TAP error bars = 0.8).}
  \label{Histogramm_11_03_29}
\end{minipage}
\begin{minipage}[]{0.45\textwidth}
  \centering
  \includegraphics[height=0.34\textheight, angle=270]{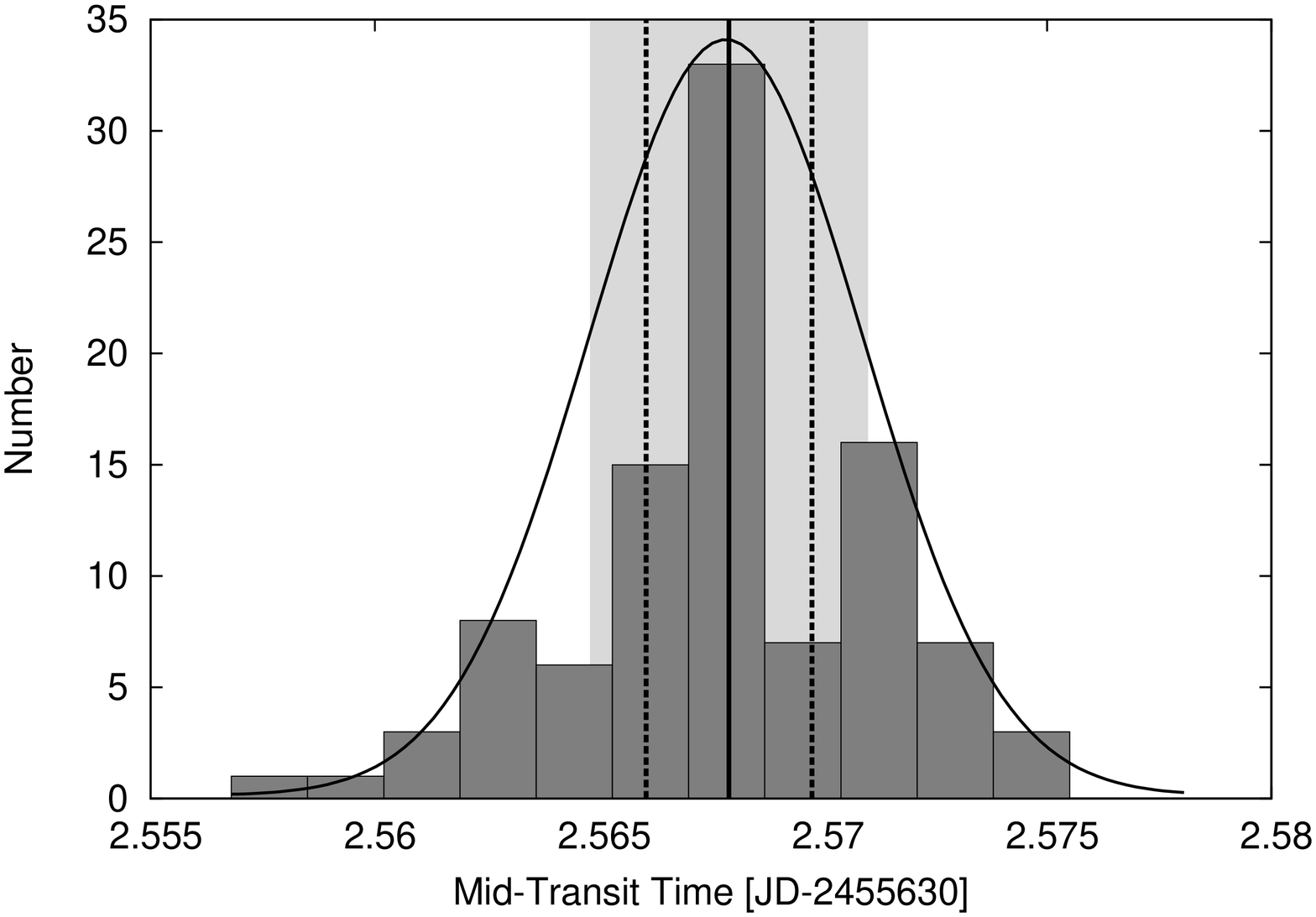}
  \caption{Same as Fig.~\ref{Histogramm_11_03_29} but for the transit at epoch 521 (2011 March 11). Given the smaller number of data points in the initial LC only 100 LCs could be created and analysed. The TAP error bars clearly underestimate the width of the transit time distribution (ratio between distribution width and TAP error bars = 1.6).}
  \label{Histogramm_11_03_11}
\end{minipage}
\end{figure}

\subsection{Timing error estimates}

The observations at the University Observatory Jena and the YETI network yield 13 transit times. Looking at the 19 individual LCs revealed some problems with several of the observations.\\ The transit from 2011 March 2 was observed with three different telescopes but all of them only yielded a partial LC due to bad weather and technical problems. Since the egress is missing in all three cases, the shape of the transit cannot be determined properly. We excluded this data point from any further timing analysis. \\ The data point from 2011 March 20 consist of three individual transits that were modelled simultaneously and therefore has small error bars. But looking at the individual transits reveals that the transit time from the Calar Alto observations differ by $\sim$10\,min from the other two LCs. Since the reason for this difference stayed unclear (most likely the bad conditions like thin clouds, fog, and full moon during the observation) we excluded the Calar Alto LC and modelled the remaining two simultaneously. \\ The LCs from 2011 February 12 and 2011 March 11 suffer from outliers and bad weather conditions in the egress, respectively. In both cases we realized significant changes in the resulting transit times for different LC treatments like removing outliers or binning. Since these effects should be accounted for in the error budget we run a test to check the robustness of the transit times. From the original (raw) LC we first created a set of 100-200 LCs (depending on the initial number of points in the LCs) by removing data points (every second, third, etc.), sigma clipping, using different binning factors, different artificial comparison stars, and removing trends. Then we modelled all LCs, determined the transit time and compiled a histogram. The width of the transit time distribution (represented by a Gauss function) gave the final timing error which were then compared to the outcome of \begin{scriptsize}TAP\end{scriptsize}. In most cases we could reproduce the uncertainties from the wavelet-based MCMC analysis (ratio between distribution width and \begin{scriptsize}TAP\end{scriptsize} error bars $\sim$\,0.8\,--\,1.2) meaning the MCMC produces very robust error bars even for LCs with a very low quality. One example for a good agreement of the \begin{scriptsize}TAP\end{scriptsize} error bars and the distribution width is given in Fig.~\ref{Histogramm_11_03_29}. Only for two LCs (2011 March 11 and 2011 April 7) the \begin{scriptsize}TAP\end{scriptsize} error bars are by a factor of 1.5-2 smaller than the distribution width as shown in Fig.~\ref{Histogramm_11_03_11}. In both cases either the ingress or the egress is affected by the weather conditions which makes it difficult to recover the transit shape. \\ As final uncertainties for the transit times, we always adopted the maximum of the \begin{scriptsize}TAP\end{scriptsize} error bars and the distribution width.

\subsection{Transit ephemeris}

To extend the observational baseline, we included the data point of \citet{2009MNRAS.392.1532J} to our timing analysis. Note, that this is not a single transit observation but the original published mid-transit time at epoch zero computed from many individual transits. Another transit observed with the University of Hawaii 2.2\,m (UH.2.2\,m) telescope on Mauna Kea is published in \citet{2009PASP..121.1104J}. The mid-transit time for this observation was determined like for our measurements as described in Section \ref{Transit_times}. With these altogether 14 mid-transit times (the partial transit at epoch 517 was excluded) we re-calculated the transit ephemeris using an error weighted linear fit. The result is given in equation (\ref{Elemente_WASP14}), where $E$ denotes the epoch (reduced $\chi^{2}$\,=\,0.64): 
\begin{equation}
\label{Elemente_WASP14}
\begin{array}{r@{.}lcr@{.}l}
T_{\mathrm{c[BJD_{TDB}]}}(E)=(2454463 & 57688 & + & E\cdot 2 & 2437655)\,\mathrm{d} \\
\pm0 & 00047 &  & \pm0 & 0000010
\end{array}
\end{equation}
The orbital period $P$ is 1.2\,s longer and 10 times more precise than the one given in \citet{2009MNRAS.392.1532J} but in agreement with \citet{2012MNRAS.426.1291S} and \citet{2013ApJ...779....5B}.

\subsection{Transit Timing Variations}

\begin{figure}
  \centering
  \includegraphics[width=0.33\textwidth,angle=270]{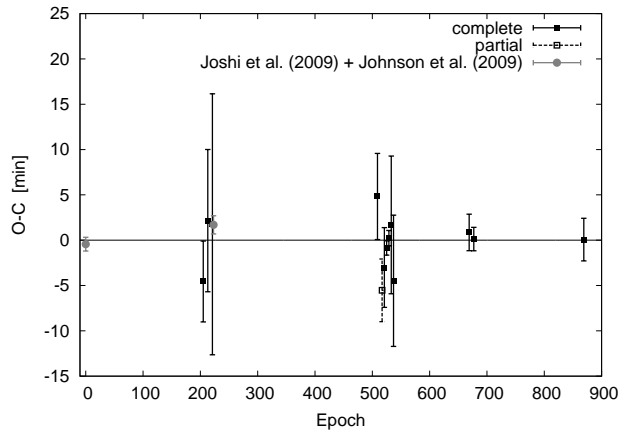}
  \caption{The O--C-diagram of WASP-14\,b. The black filled and open (with dashed error bars) symbols denote the complete and the partial transits, respectively. Note that the transit at epoch 669 (2012 February 6) is considered here as `complete' since the ingress as well as the egress, and therefore the transit shape, are intact. The data points from \citet{2009MNRAS.392.1532J} and \citet{2009PASP..121.1104J} are shown in grey. The solid line represents the updated ephemeris given in equation (\ref{Elemente_WASP14}). }
  \label{O_C_2014}
\end{figure}

\begin{figure}
  \centering
  \includegraphics[width=0.32\textwidth,angle=270]{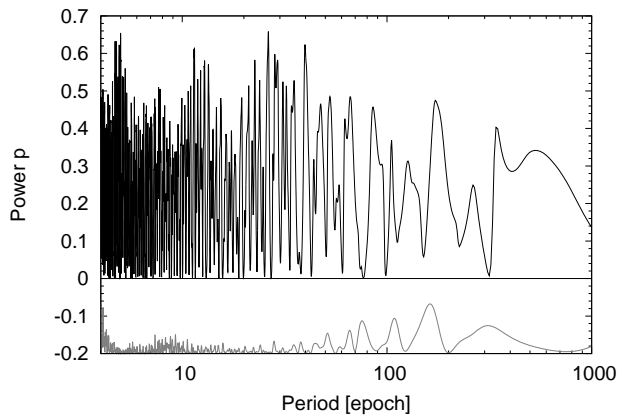}
  \caption{Generalized Lomb--Scargle periodogram (top panel) and window function (bottom panel) for the O--C diagram of WASP-14\,b. The highest peak with a period of $P_{\mathrm{TTV}}$\,=\,26.33\,$\pm$\,0.09\,epochs at a power of 0.66 shows a FAP of 44.7\%.}
  \label{GLS_plot}
\end{figure}

\begin{figure}
  \centering
  \includegraphics[width=0.45\textwidth]{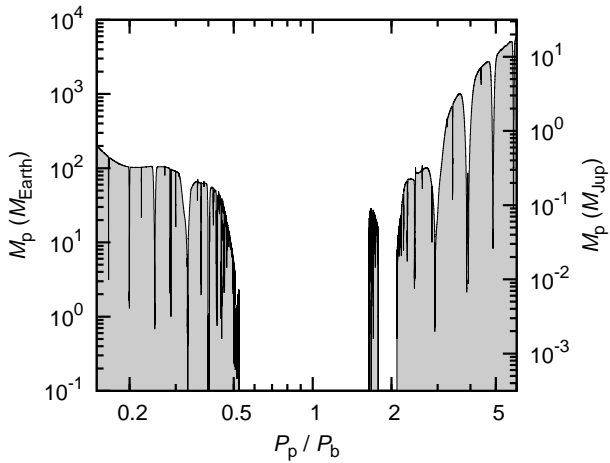}
  \caption{Upper mass limit for a fictitious additional planet in the WASP-14 system, based on the timing data set, as a function of the orbital period of that planet, $P_{\mathrm{p}}$. The greyed area shows unexplored configurations because they are below a detection threshold of the timing data set.}
  \label{fig-limit}
\end{figure}

If we subtract the predicted transit times (calculated with the refined ephemeris) from the observed value, we obtain the O--C. The results for the mid-transit times and the O--C are listed in Table~\ref{WASP14_Transit_Times}. Fig.~\ref{O_C_2014} shows the O--C diagram where the black solid line represents the refined ephemeris. To search for a periodicity in the O--C diagram we computed the generalized Lomb--Scargle periodogram \citep[\begin{scriptsize}GLS\end{scriptsize};][]{2009A&A...496..577Z}. The periodogram is shown in Fig.~\ref{GLS_plot} where the highest peak ($P_{\mathrm{TTV}}$\,=\,26.33\,$\pm$\,0.09\,epochs, power of 0.66) corresponds to a False-Alarm probability (FAP) of 44.7\%. Within our data set we could not detect any evidence for TTV.\\ Residuals in transit timing allow us to place constraints on the properties of any hypothetical additional planet in the system. A set of synthetic O--C diagrams for WASP-14\,b in the presence of a fictitious perturbing planet was generated with the \begin{scriptsize}MERCURY\end{scriptsize} 6 package \citep{1999MNRAS.304..793C} and the implemented Bulirsch--Stoer integrator. The mass of this second planet was set at 0.5, 1, 5, 10, 50, 100, and 500 $M_{\mathrm{Earth}}$ (Earth masses). The initial semimajor axis was iterated from 0.01 to 0.12 au with a step of $2\times10^{-6}$ au. The system was assumed to be coplanar. For WASP-14\,b, the initial argument of periastron, $\omega_{\mathrm{b}}$, and orbital eccentricity, $e_{\mathrm{b}}$ were taken from Blecic et al. (2013), and the mean anomaly was set at $0^{\circ}$. For the fictitious planet, its argument of periastron was set at the value equal to $\omega_{\mathrm{b}}$ at the beginning of each simulation. Calculations were done for three values of the orbital eccentricity of the fictitious planet, $e_{\mathrm{p}}$, set to 0.0, 0.1, and 0.2. The initial mean anomaly was shifted by $180^{\circ}$. The integration for each planetary configuration covered 2250\,d (i.e. 1000 orbit of WASP-14\,b -- the time span of transit timing observations). We calculated the rms of the signal at $P_{\mathrm{TTV}}$ as rms$_{\mathrm{TTV}}=A_{\mathrm{max}} / \sqrt{2}$, where $A_{\mathrm{max}}$ is an amplitude of this signal. We derived rms$_{\mathrm{TTV}}=37\,\mathrm{s}$. Analogously, rms of the residuals from a linear ephemeris was calculated for each set of simulated observations. Then, for each orbital distance, we determined the range of planet masses in which the value of rms$_{\mathrm{TTV}}$ fell. An upper mass of the fictitious planet at the detection limit was found by linear interpolation for masses below 500 $M_{\mathrm{Earth}}$. If the TTV sinal was found to be generated by a more massive body, the limiting mass was extrapolated using a linear trend as fitted to 100 and 500 $M_{\mathrm{Earth}}$. \\ Exemplary results for $e_{\rm{p}}=0.1$ are illustrated in Fig.~\ref{fig-limit}. The timing technique allows us to probe the Earth-mass regime close to MMRs. Most of orbits located between the inner 1:2 and outer 2:1 orbital period commensurabilities, were found to be highly unstable and planetary close encounters or planet ejections occurred during the relatively short time of integration. Thus, planetary configurations which are placed in this space are highly unlikely to exist.

\begin{table*}
\centering
\caption{Transit times for all observed transits of WASP-14\,b including the publicly available transits. The O--C was calculated with the ephemeris given in equation (\ref{Elemente_WASP14}).}
\label{WASP14_Transit_Times}
\renewcommand{\arraystretch}{1.1}
\begin{tabular}{ccr@{\,$\pm$\,}lr@{\,$\pm$\,}lc}
\hline \hline
 Date & Epoch & \multicolumn{2}{c}{$T_{\mathrm{c}}$ (BJD)} & \multicolumn{2}{c}{O--C  (min)} & Ref. \\ \hline \hline
            & 0    & 2454463.57657 & 0.00053                & -0.44 & 0.76 & \citet{2009MNRAS.392.1532J} \\
2009 Apr. 1  & 205  & 2454923.54564 & $^{0.00310}_{0.00310}$ & -4.55 & $^{4.46}_{4.46}$ & This work \\
2009 Apr. 19 & 213  & 2454941.50043 & $^{0.00545}_{0.00545}$ & 2.16  & $^{7.85}_{7.85}$ & This work \\
2009 May 7  & 221  & 2454959.45027 & $^{0.01000}_{0.00760}$ & 1.75  & $^{14.40}_{10.94}$ & This work \\
            & 223$^{a}$  & 2454963.93776 & $^{0.00069}_{0.00070}$ & 1.69  & $^{0.99}_{1.01}$ & \citet{2009PASP..121.1104J} \\ 
2011 Feb. 12 & 509  & 2455605.65997 & $^{0.00330}_{0.00310}$ & 4.83  & $^{4.75}_{4.46}$ & This work \\
(2011 Mar. 2  & 517$^{b}$ & 2455623.59980 & $^{0.00240}_{0.00200}$ & -5.53 & $^{3.46}_{2.88}$ & This work)* \\
2011 Mar. 11 & 521  & 2455632.57247 & $^{0.00306}_{0.00306}$ & -3.01 & $^{4.40}_{4.40}$ & This work \\
2011 Mar. 20 & 525$^{b}$ & 2455641.54831 & $^{0.00055}_{0.00055}$ & -0.86 & $^{0.79}_{0.79}$ & This work \\
2011 Mar. 29 & 529$^{b}$ & 2455650.52899 & $^{0.00058}_{0.00058}$ & 0.24  & $^{0.84}_{0.84}$ & This work \\
2011 Apr. 7  & 533  & 2455659.50506 & $^{0.00528}_{0.00528}$ & 1.69  & $^{7.60}_{7.60}$ & This work \\
2011 Apr. 16 & 537  & 2455668.47584 & $^{0.00503}_{0.00503}$ & -4.48 & $^{7.24}_{7.24}$ & This work \\
2012 Feb. 6  & 669$^{b}$ & 2455964.65659 & $^{0.00140}_{0.00140}$ & 0.86  & $^{2.02}_{2.02}$ & This work \\
2012 Feb. 24 & 677  & 2455982.60621 & $^{0.00090}_{0.00090}$ & 0.13  & $^{1.29}_{1.29}$ & This work \\
2013 Apr. 30 & 869  & 2456413.40914 & $^{0.00163}_{0.00163}$ & 0.07  & $^{2.35}_{2.35}$ & This work \\
\hline \hline
\end{tabular}
\\
$^{\ast}$Data point was excluded from timing analysis.\\
$^{a}$Transit was re-analysed with \begin{scriptsize}TAP\end{scriptsize}.\\
$^{b}$The LCs have been combined during the MCMC analysis to improve timing precision.
\end{table*}

\section{Conclusions}

\begin{figure}
  \centering
  \includegraphics[width=0.33\textwidth,angle=270]{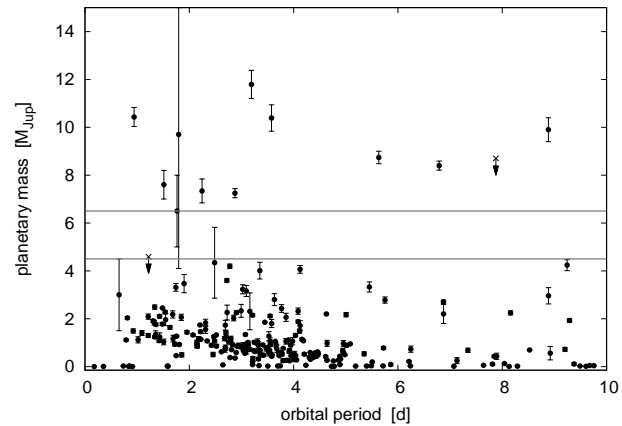}
  \caption{The period-mass diagram for close-in exoplanets ($P\,<\,10$\,d and $M\,<\,15\,M_{\mathrm{Jup}}$). Only 12 planets are more massive than 6.5\,$M_{\mathrm{Jup}}$. No planets were found in he region between 4.5 and 6.5\,$M_{\mathrm{Jup}}$ (except WASP-33\,b which is just an upper mass limit $M\,<\,4.59\,M_{\mathrm{Jup}}$, upper limits are marked with an arrow)}
  \label{Transit_period_mass.ps}
\end{figure}

\begin{figure*}
 \centering
  \includegraphics[width=0.21\textwidth, angle=270]{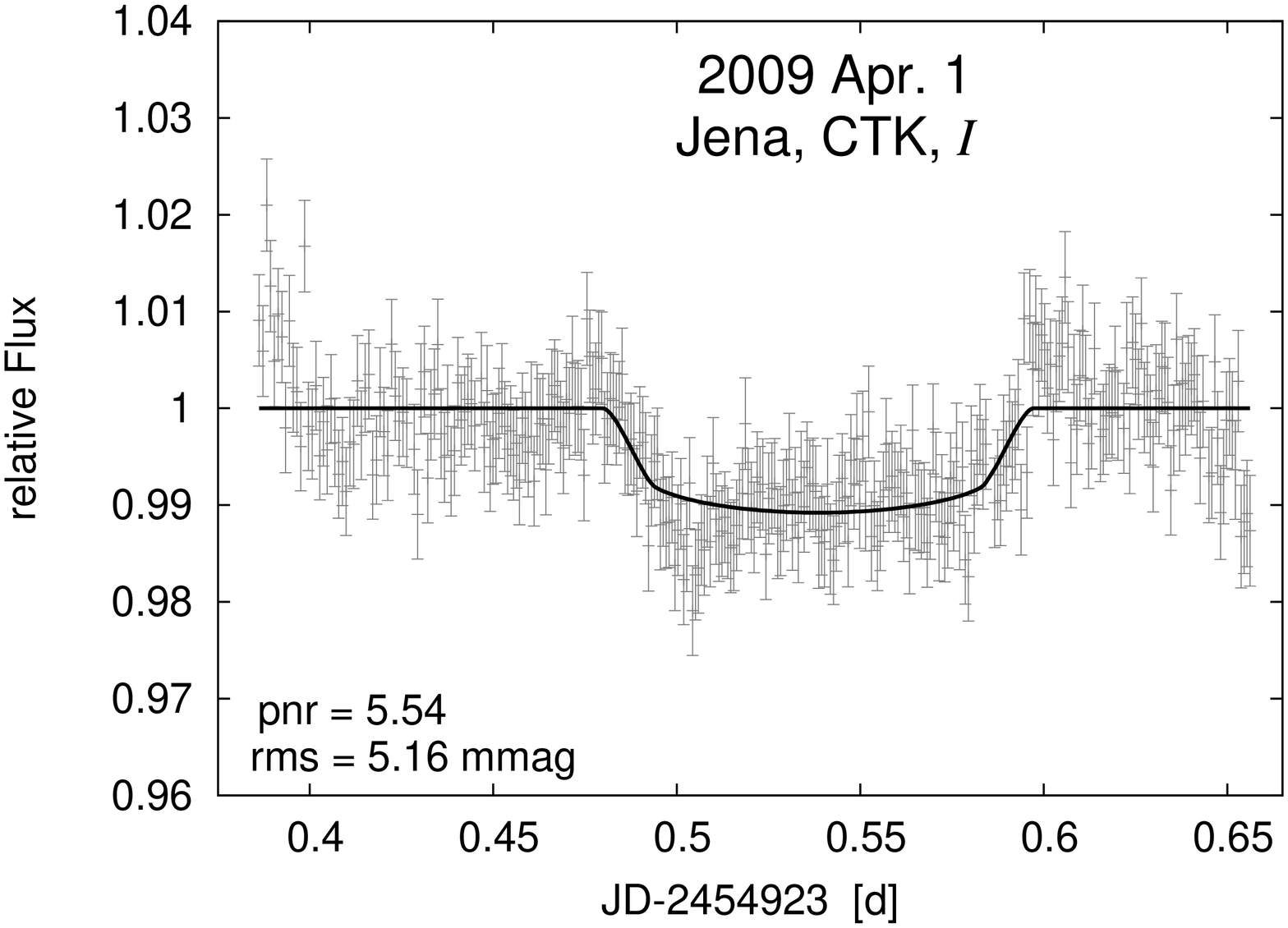}
  \includegraphics[width=0.21\textwidth, angle=270]{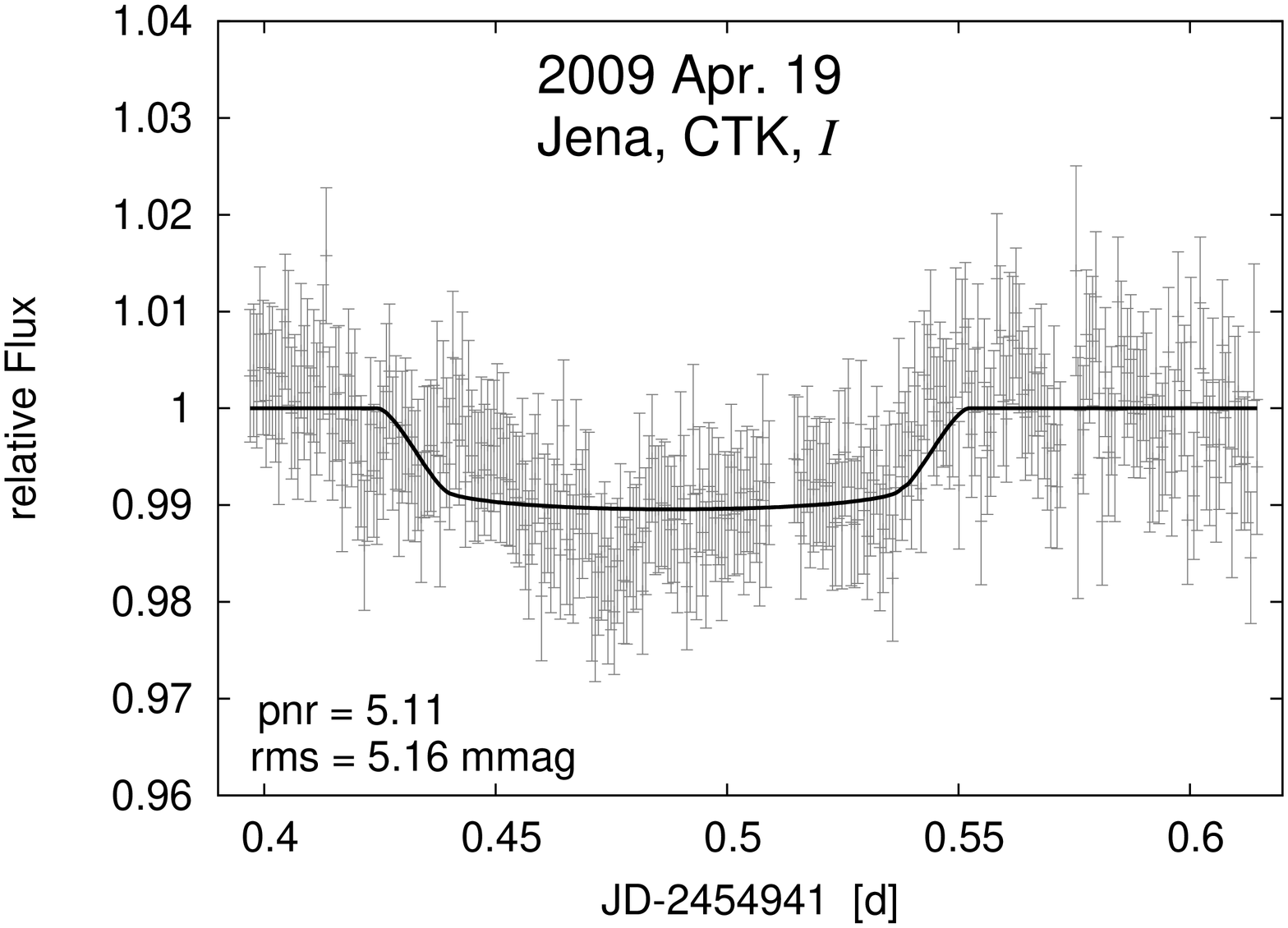}
  \includegraphics[width=0.21\textwidth, angle=270]{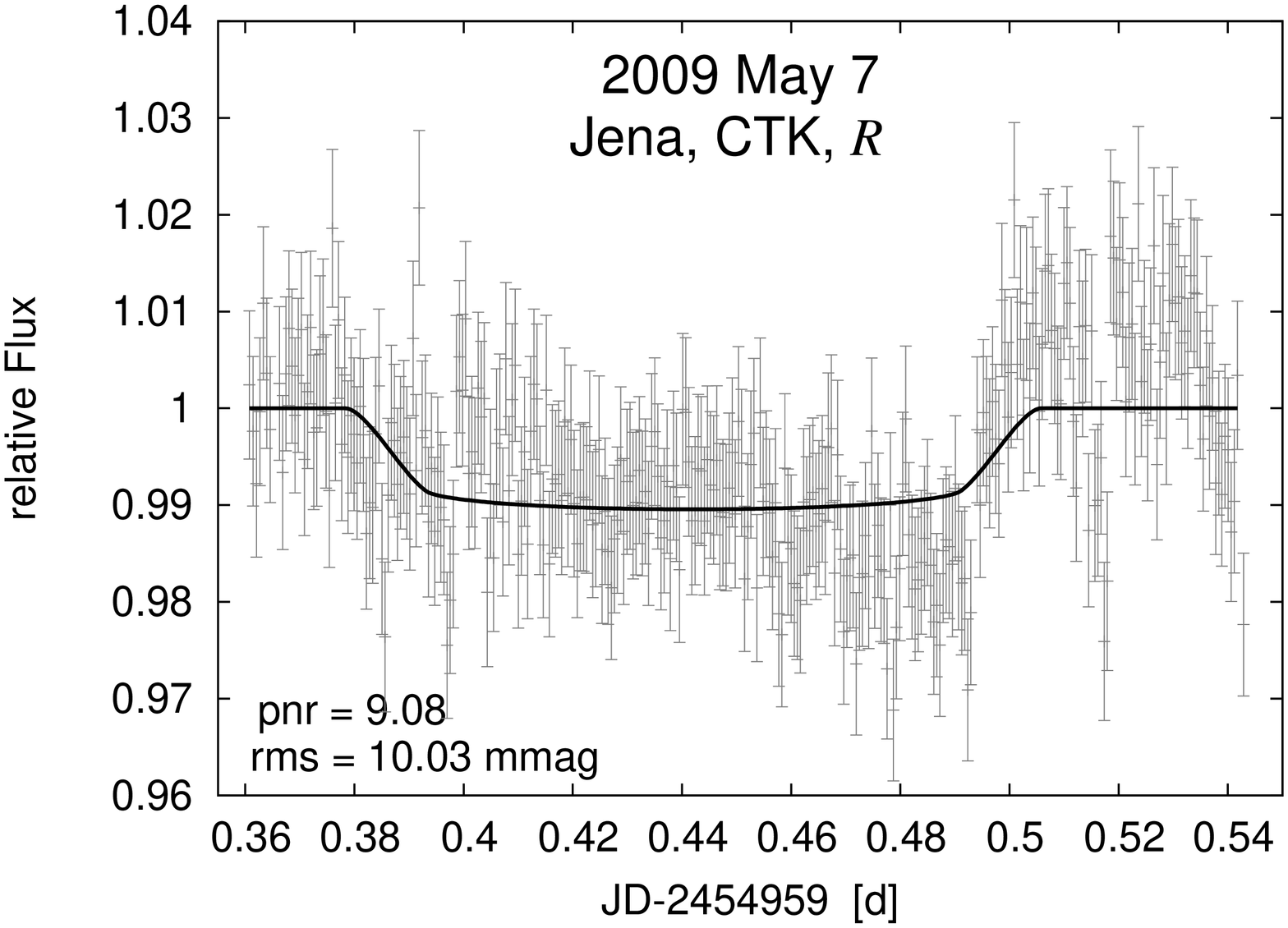}
  \includegraphics[width=0.21\textwidth, angle=270]{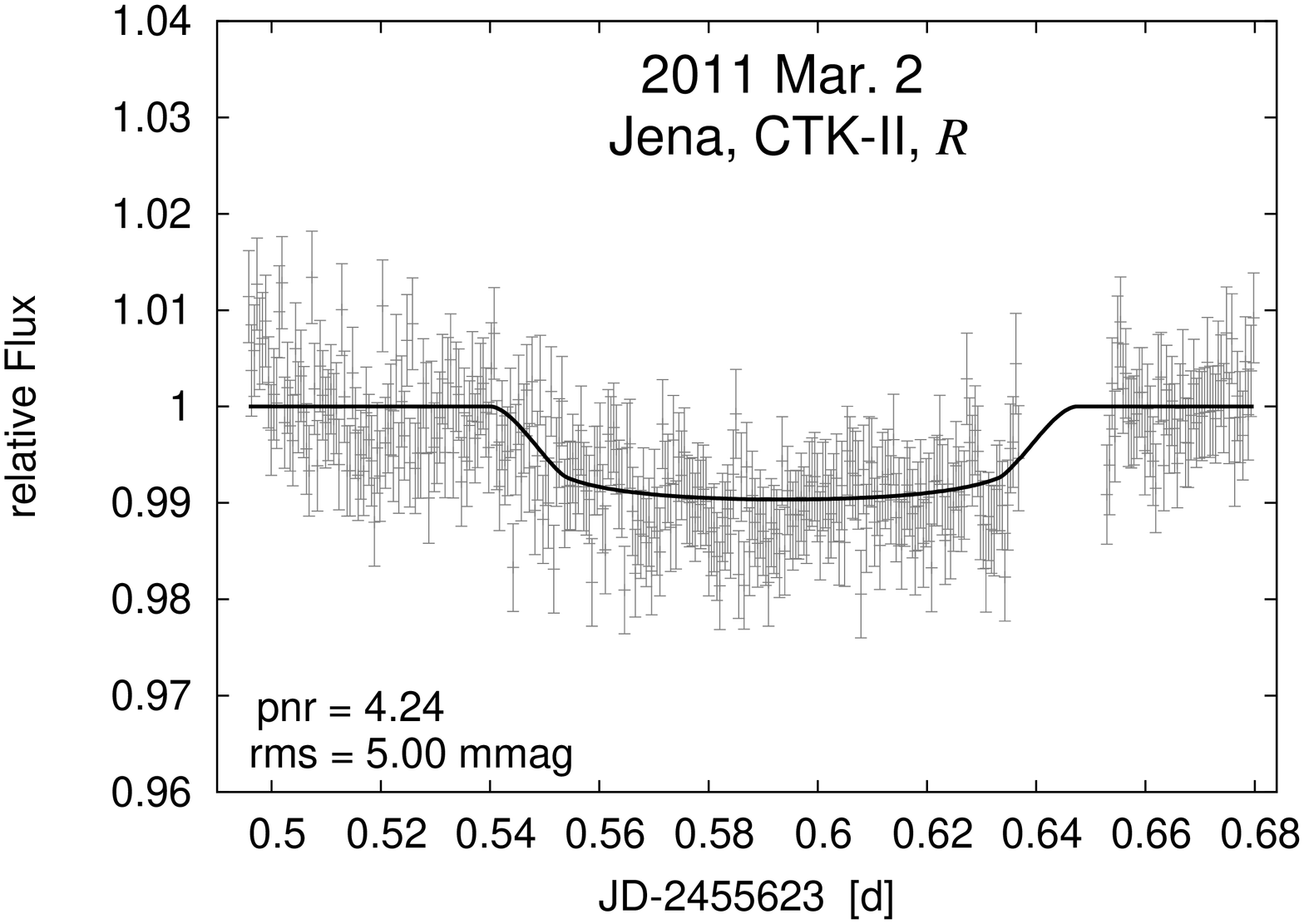}
  \includegraphics[width=0.21\textwidth, angle=270]{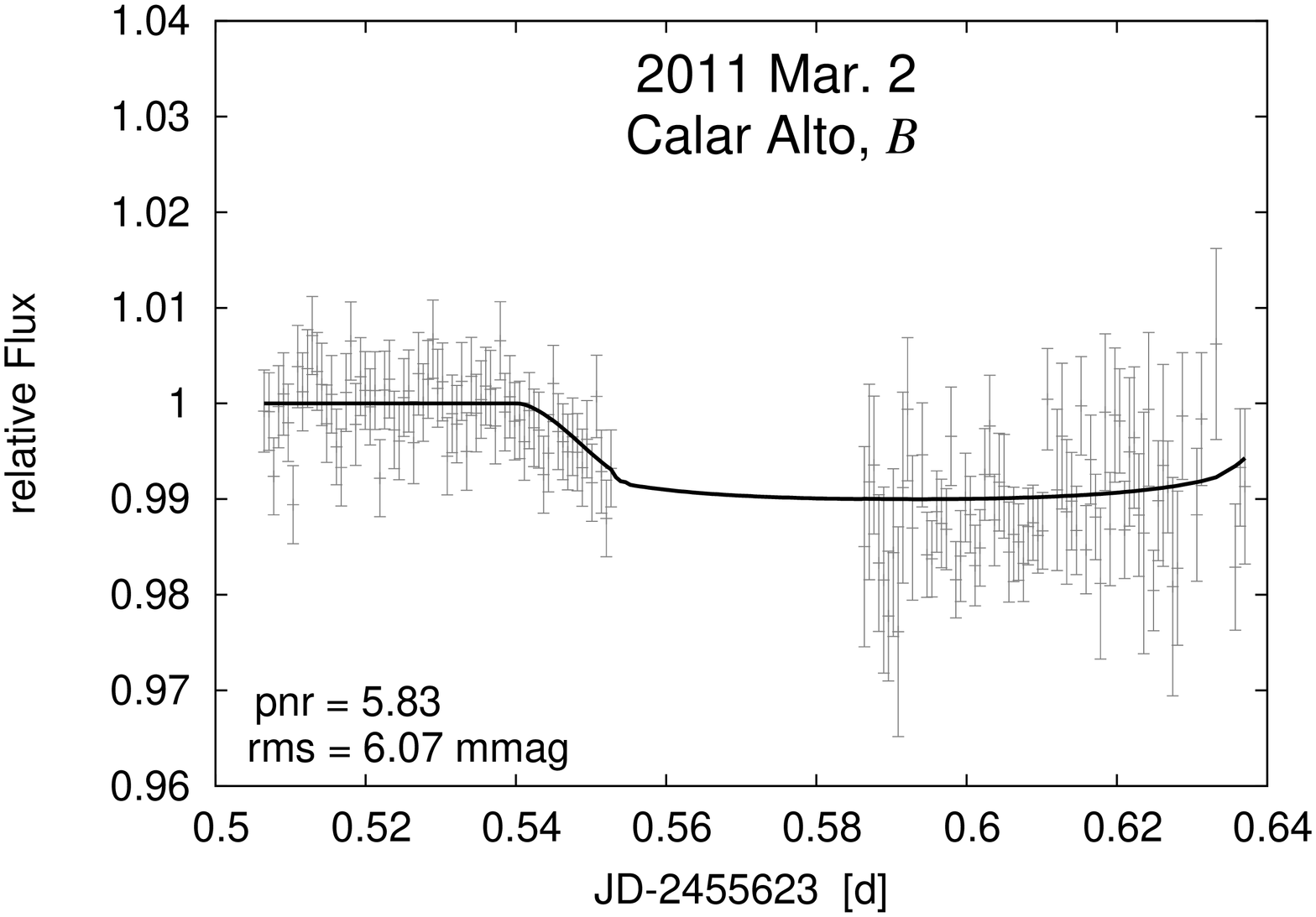}
  \includegraphics[width=0.21\textwidth, angle=270]{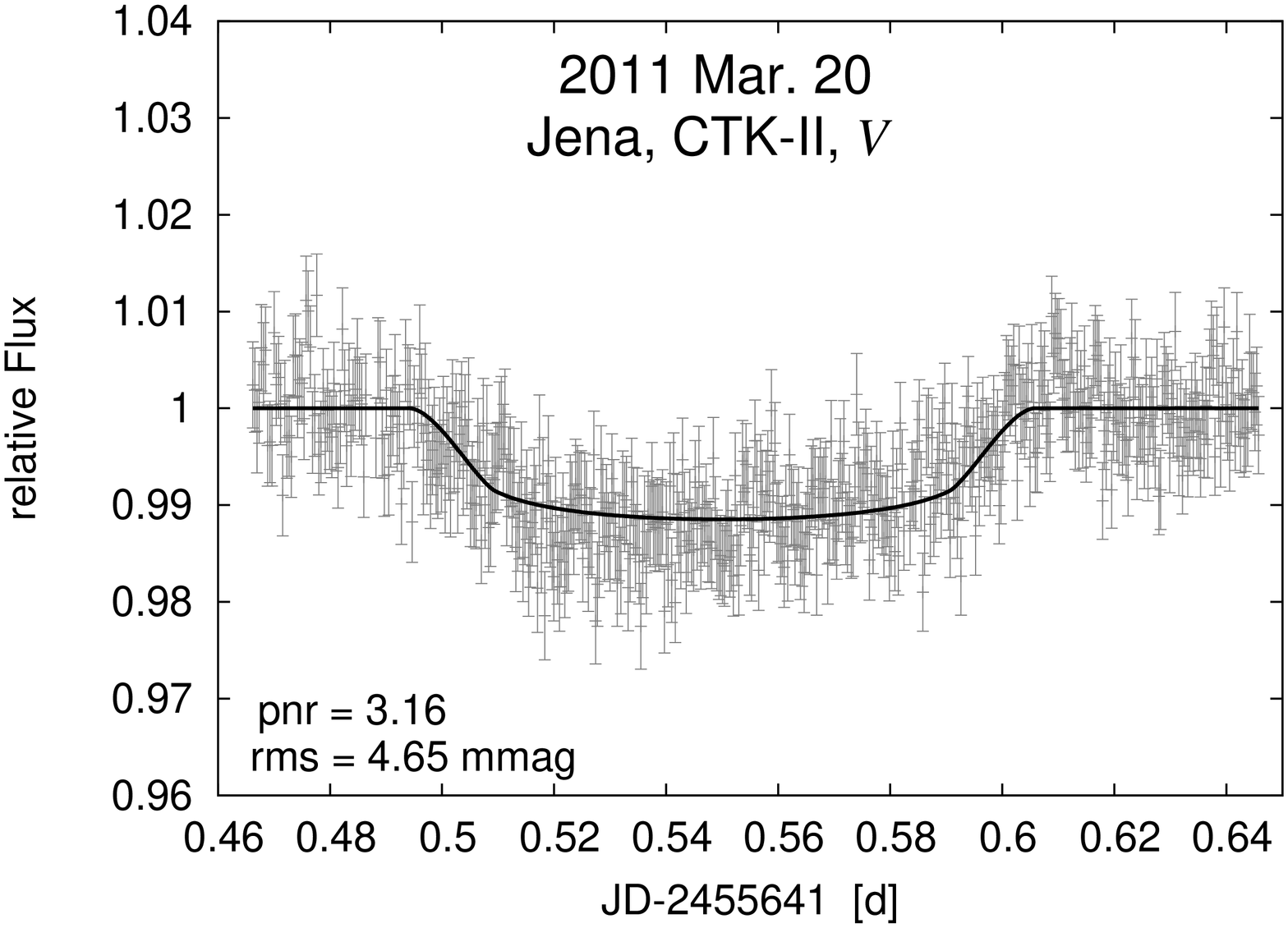}
  \includegraphics[width=0.21\textwidth, angle=270]{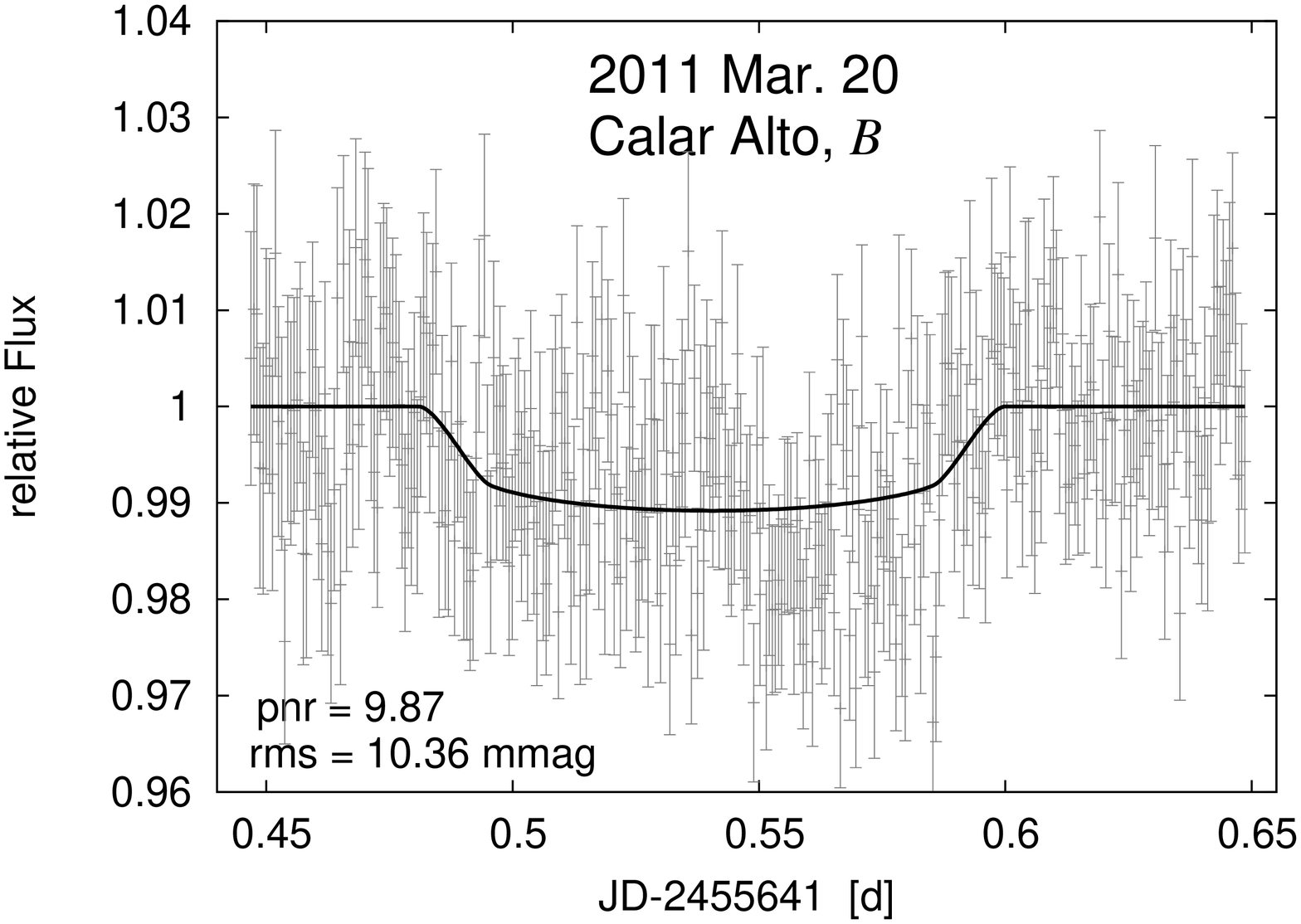}
  \includegraphics[width=0.21\textwidth, angle=270]{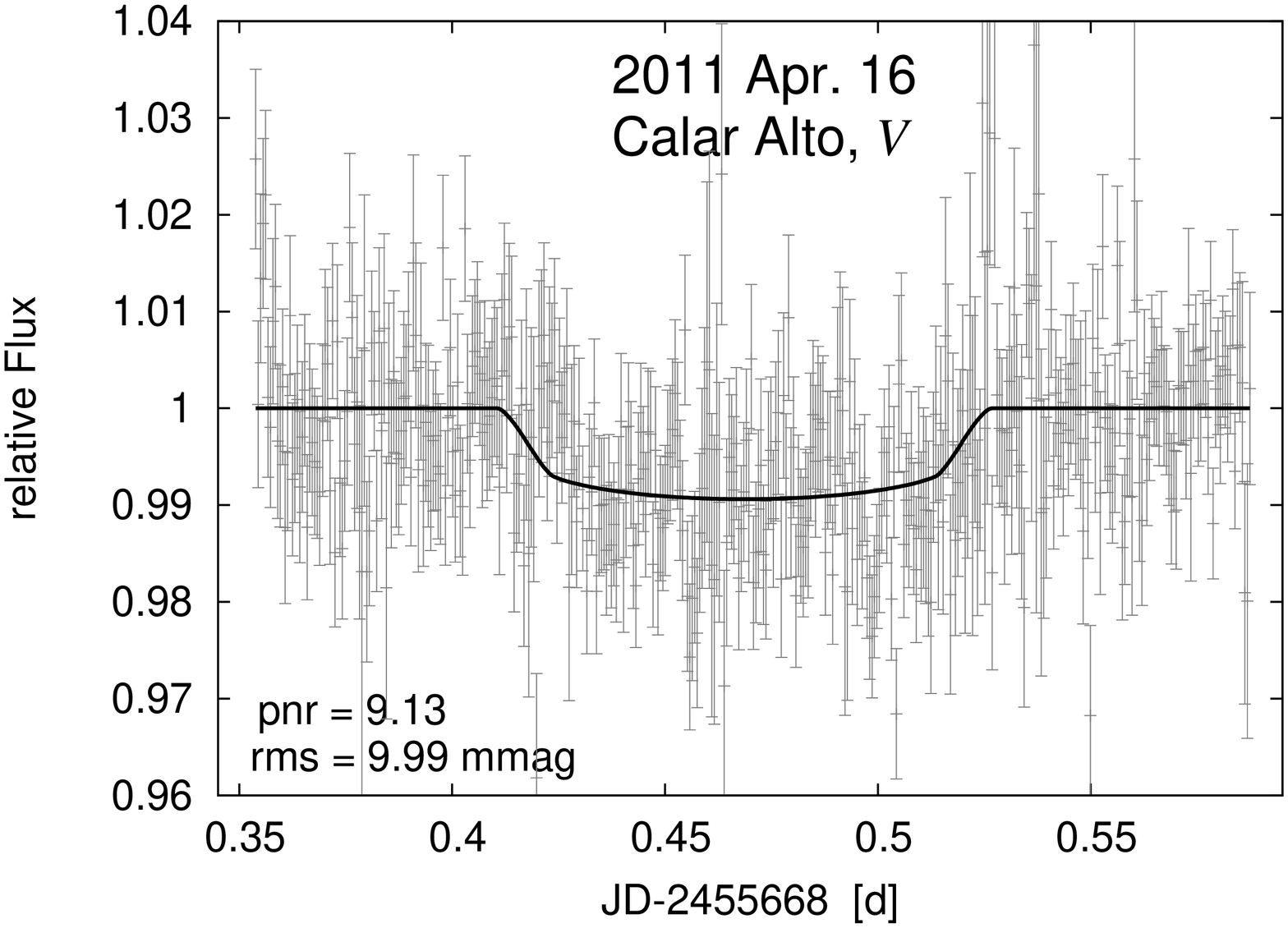}
  \includegraphics[width=0.21\textwidth, angle=270]{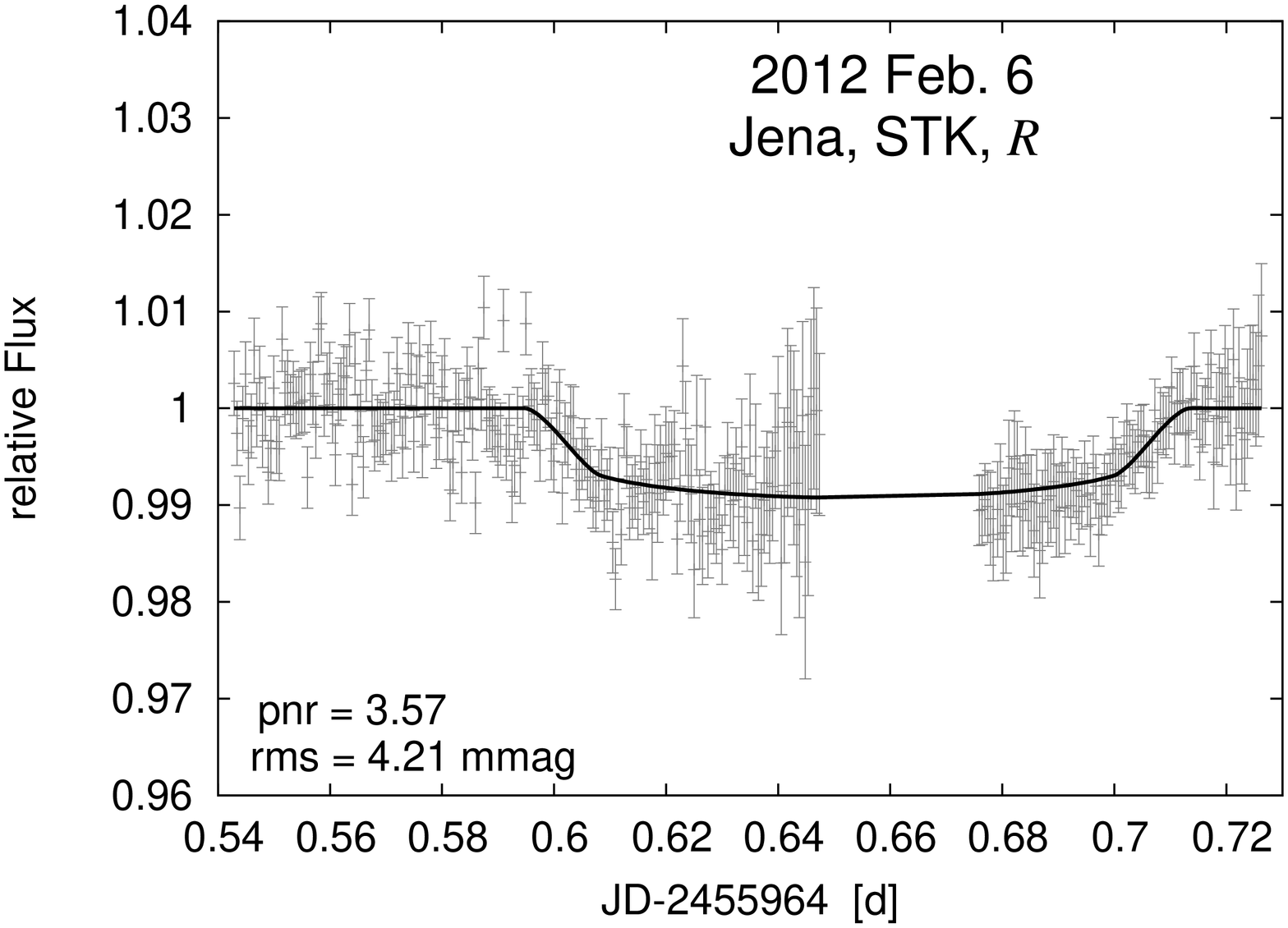}
  \includegraphics[width=0.21\textwidth, angle=270]{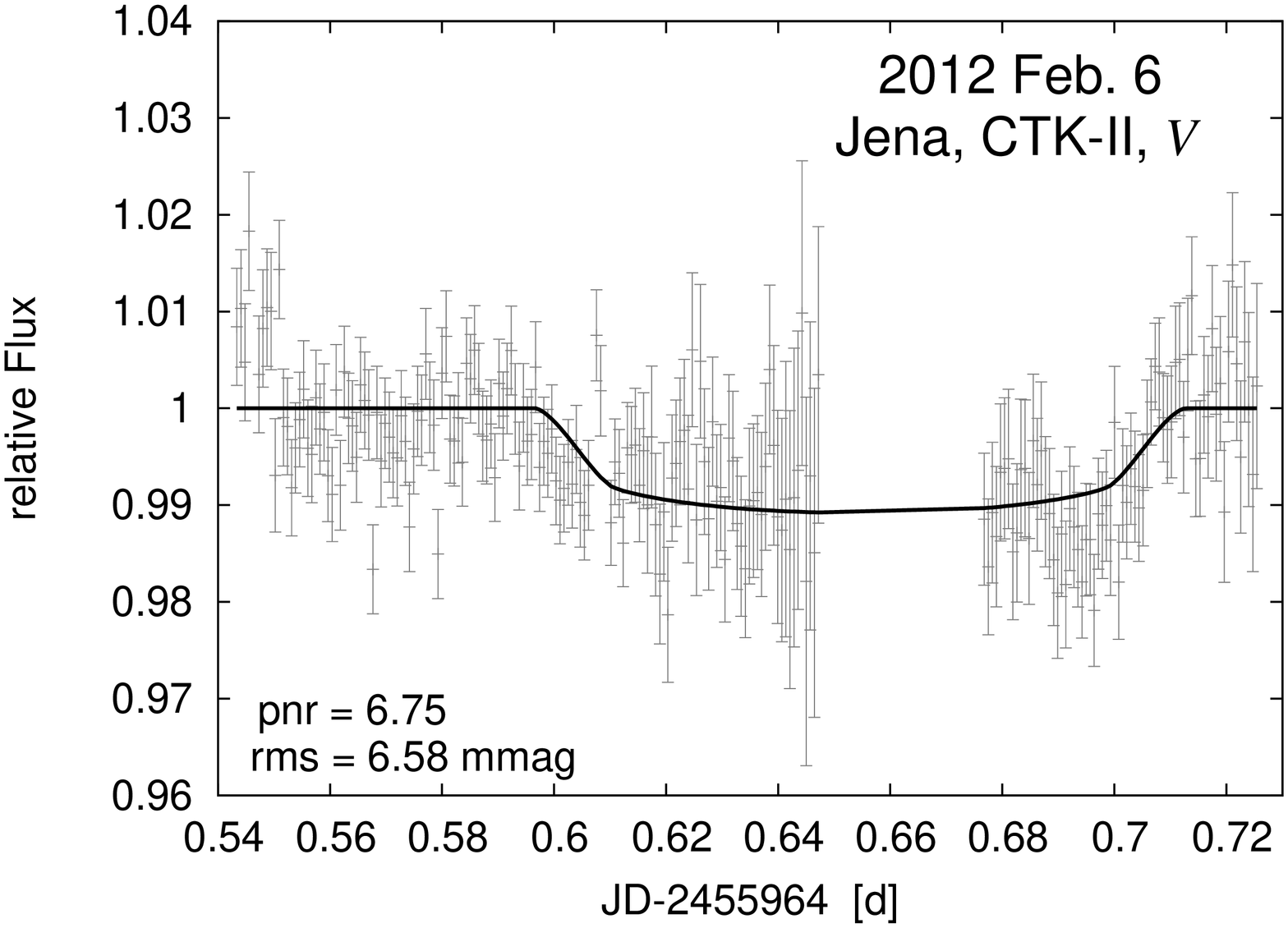}
  \caption{LCs of WASP-14\,b with an rms$\,>$\,4\,mmag. The date of observation, observatory, filter, pnr, and the rms of the fit are indicated in each individual panel}
\label{LC_Wasp14a}
\end{figure*}

\begin{figure*}
 \centering
  \includegraphics[width=0.21\textwidth, angle=270]{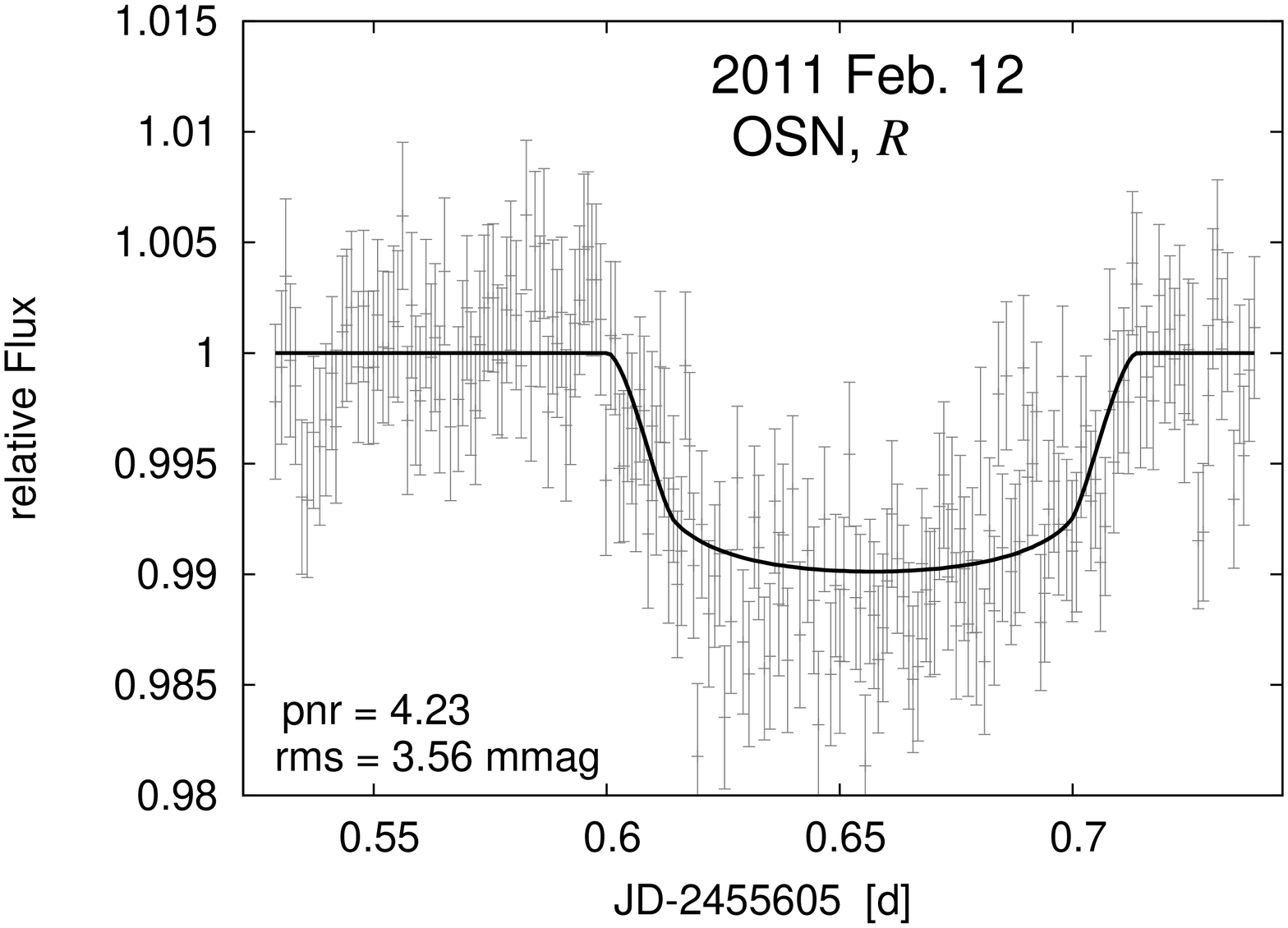}
  \includegraphics[width=0.21\textwidth, angle=270]{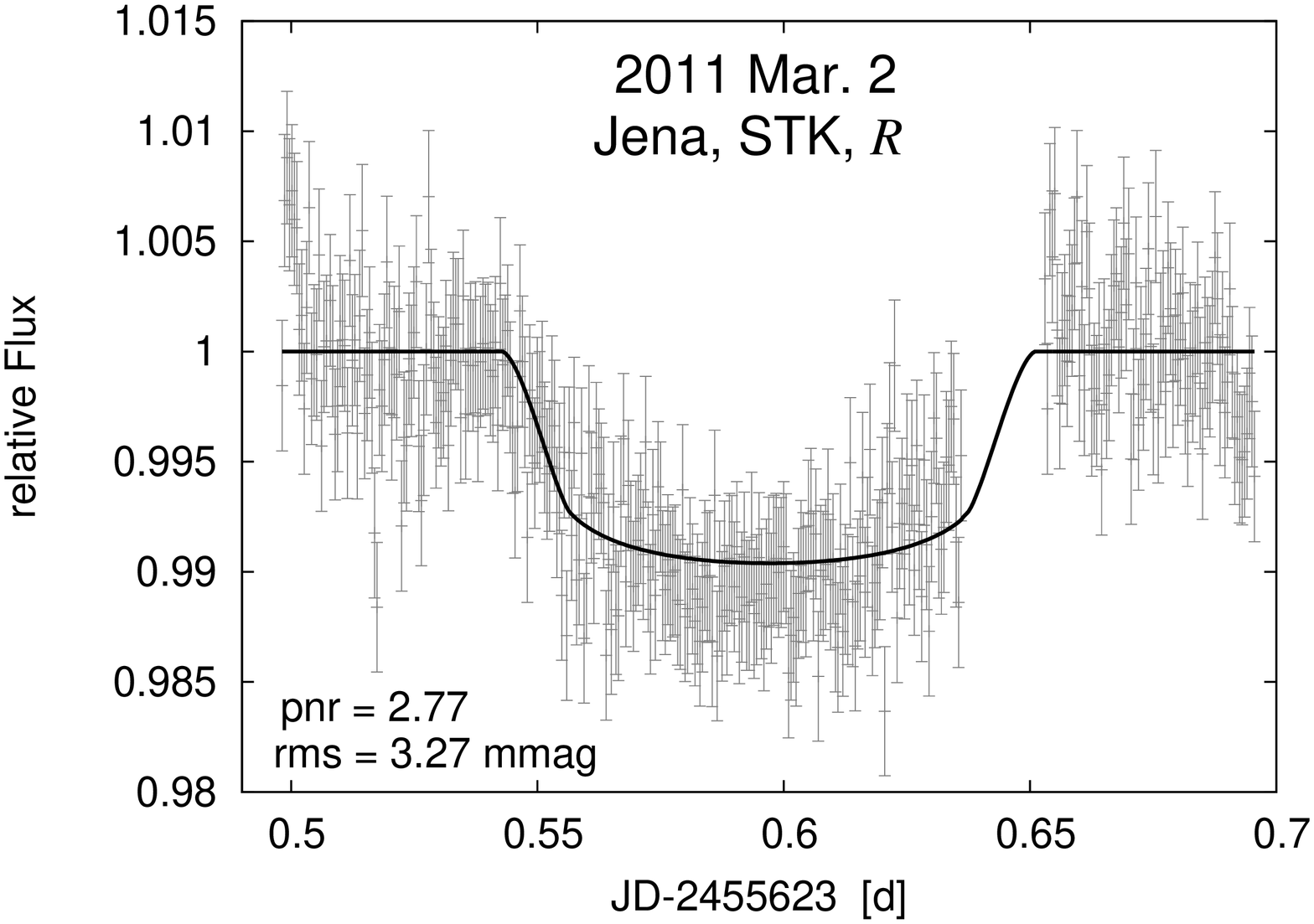}
  \includegraphics[width=0.21\textwidth, angle=270]{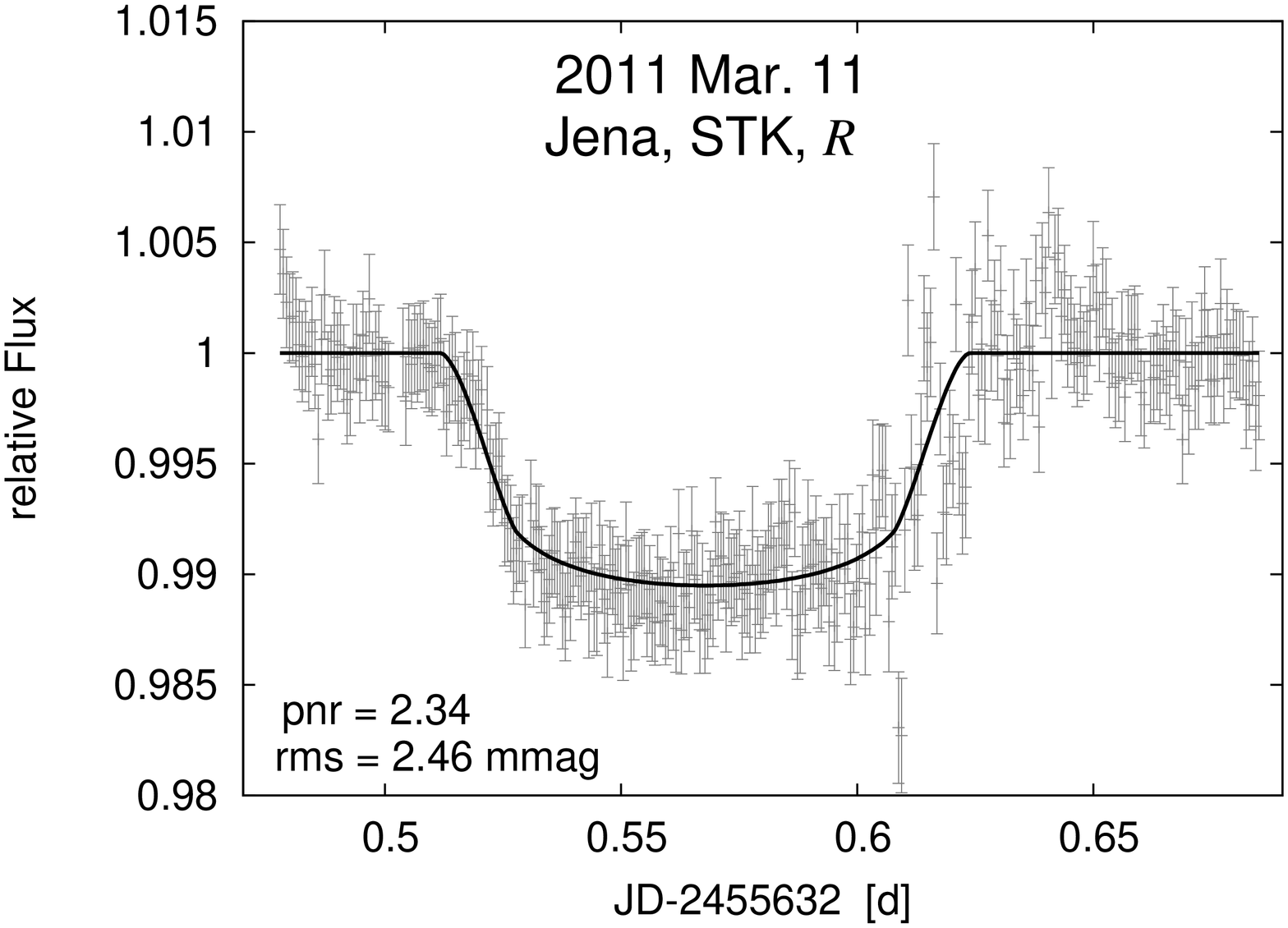}
  \includegraphics[width=0.21\textwidth, angle=270]{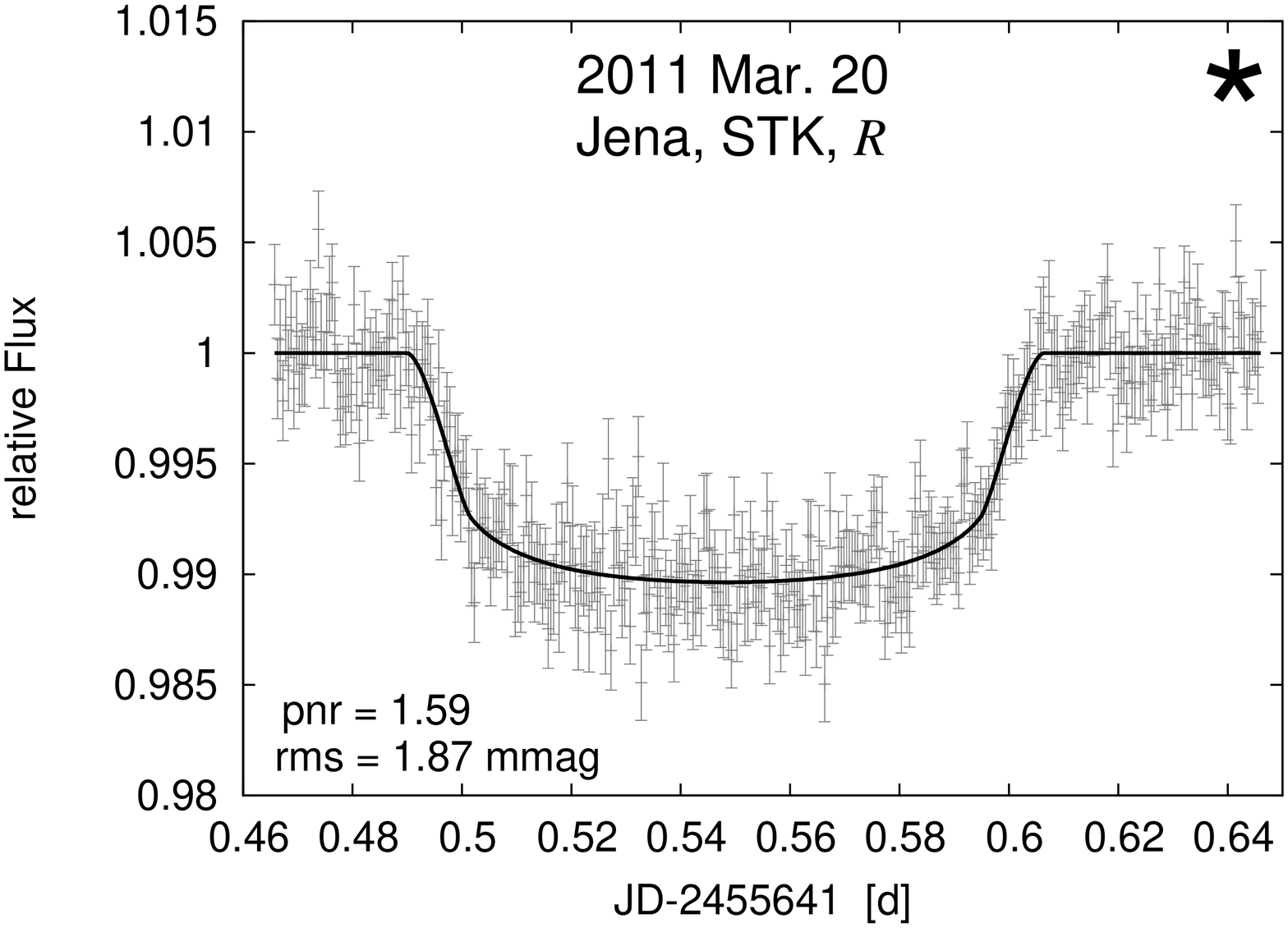}
  \includegraphics[width=0.21\textwidth, angle=270]{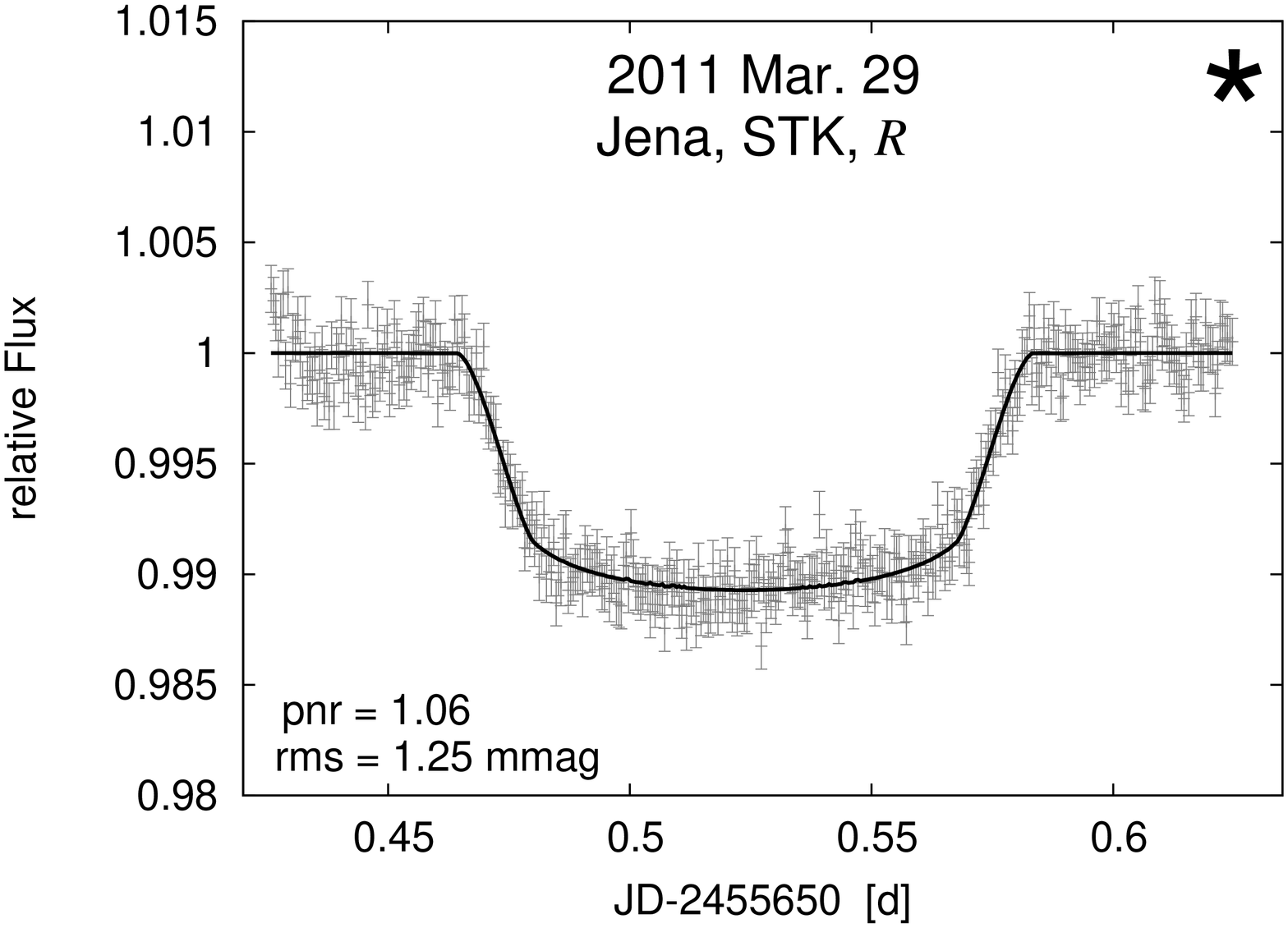}
  \includegraphics[width=0.21\textwidth, angle=270]{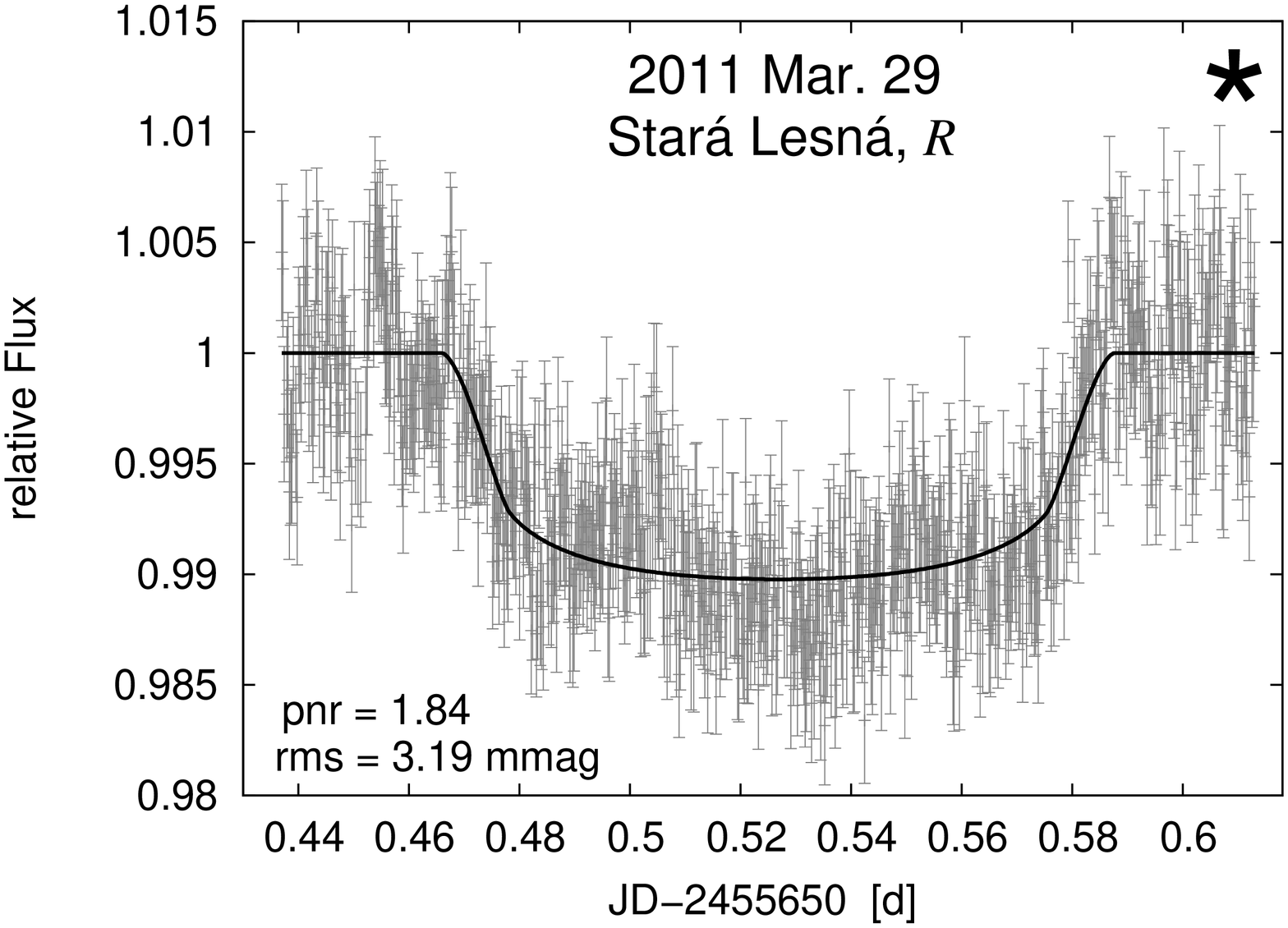}
  \includegraphics[width=0.21\textwidth, angle=270]{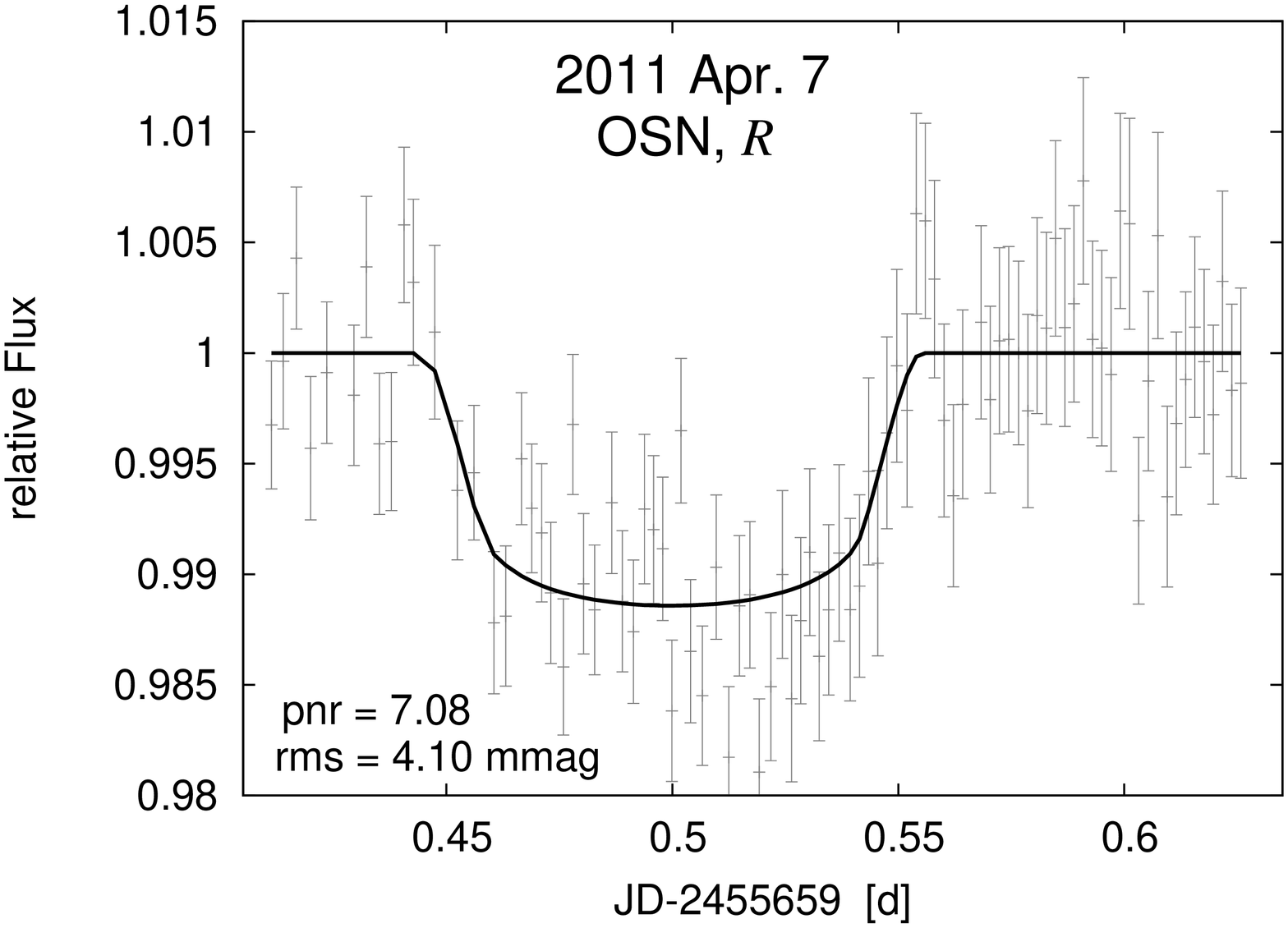}
  \includegraphics[width=0.21\textwidth, angle=270]{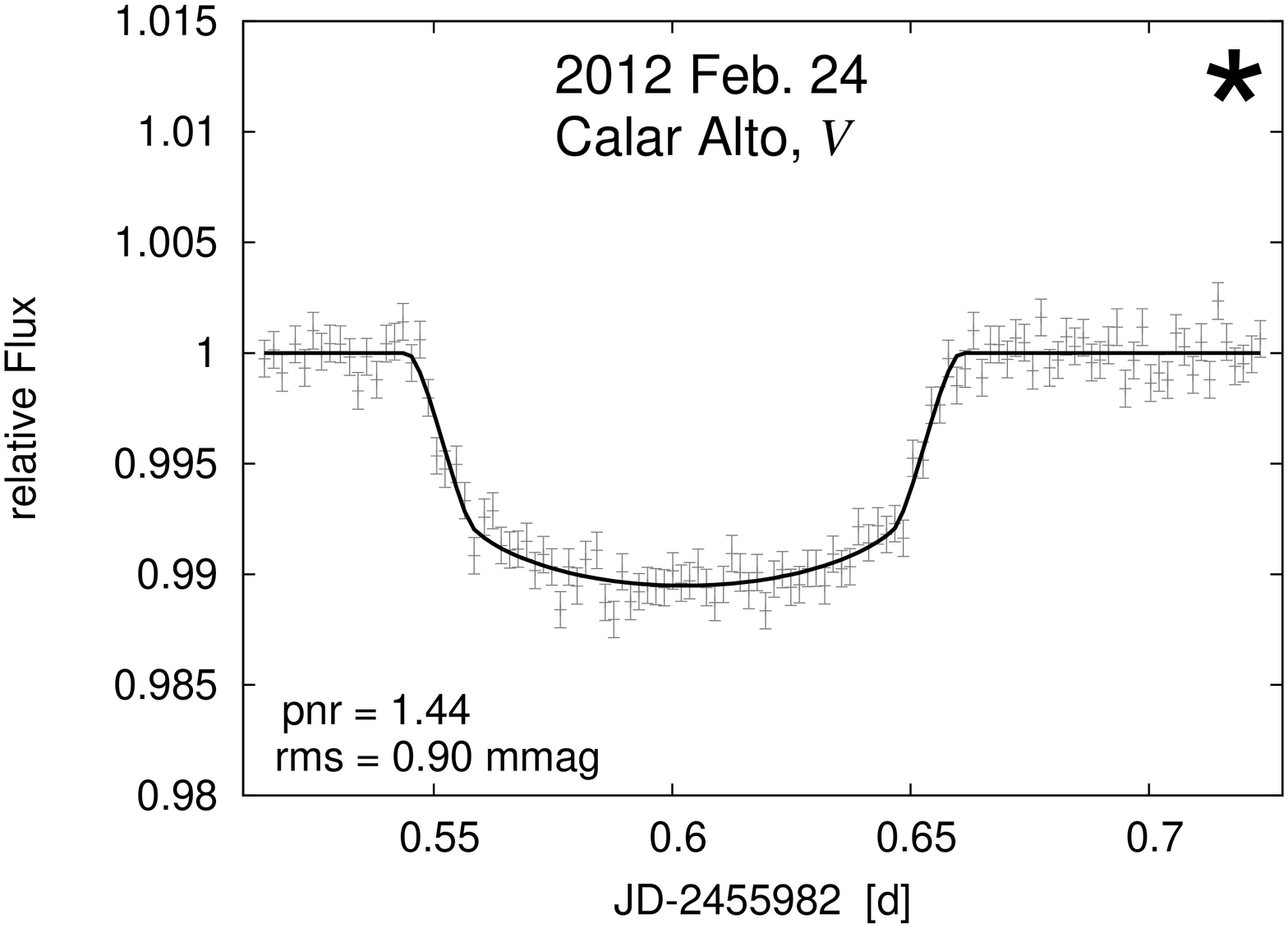}
  \includegraphics[width=0.21\textwidth, angle=270]{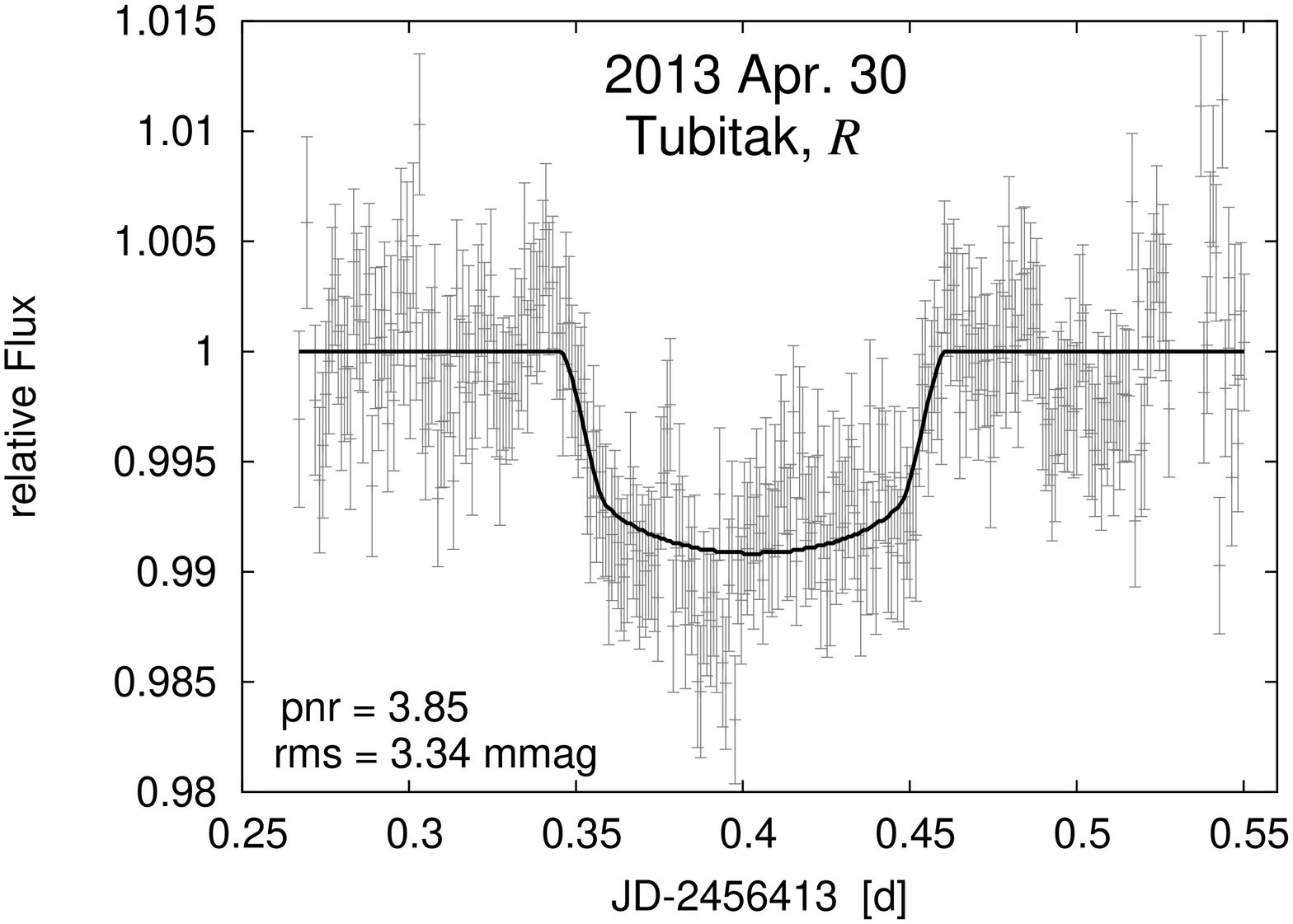}  
  \caption{The same as Fig. \ref{LC_Wasp14a} but for the higher quality LCs with an rms$\,<$\,4\,mmag. The four best quality LCs (in terms of pnr) that were used to create the template are marked with an asterisk.} 
\label{LC_Wasp14b}
\end{figure*}

WASP-14\,b is, for several reasons, a very interesting target. First, because of its very high mass it is one of the densest exoplanet with an orbital period shorter than three days. Furthermore, despite of its close-in orbit WASP-14\,b has a rather high eccentricity. Interestingly, there is a strong tendency that massive planets on close-in orbits have elliptical orbits. Approximately 58\% (7 out of 12, exoplanet.eu, 2014 November 12) of transiting exoplanets with masses greater than  $M\,=\,5\,M_{\mathrm{Jup}}$ and  periods less than 10\,d have $e\,\neq$\,0 while only 20\% (53 out of 265, exoplanet.eu, 2014 November 12) of the less massive planets show a significant eccentricity. This may indicate that there are two distinct types of exoplanets. Another hint is the distribution in the period-mass diagram for close-in exoplanets ($P\,<\,10$\,d and $M\,<\,15\,M_{\mathrm{Jup}}$) which is shown in Fig.~\ref{Transit_period_mass.ps}. Two distinct areas are clearly identified. Most of the planets have a mass below 5\,$M_{\mathrm{Jup}}$. So far, no planets were discovered in the mass range between 4.5 and 6.5\,$M_{\mathrm{Jup}}$ (The only point in this area belongs just to an upper mass limit of WASP-33\,b, $M\,<\,4.59\,M_{\mathrm{Jup}}$), while several objects were found with masses $\geq$\,6.5\,$M_{\mathrm{Jup}}$, which are often referred to as hot super-Jupiter. This clear distinction could indicate a physical difference in the planet formation and evolution. Since only a handful of hot super-Jupiter known so far, this massive area is insufficiently explored to make general statements. \\ We observed 19 LCs of WASP-14\,b with six telescopes at five different observatories within the TTV$@$YETI collaboration. Due to the weather conditions and observations done with small telescopes the LCs are of highly variable quality. All transits are shown in Figs~\ref{LC_Wasp14a} and \ref{LC_Wasp14b}, grouped according to their quality to LC with rms$\,>$\,4\,mmag and $rms\,<$\,4\,mmag, respectively. \\ From the simultaneous LC modelling of our four best LCs we could determine planetary, stellar and geometrical properties of the system. Our values are in agreement with the values in \citet{2009MNRAS.392.1532J} but differ significantly from the physical properties given in \citet{2012MNRAS.426.1291S}. This discrepancy, however, is not physical since it is only caused by the interpretation of the final results by \citet[][averaging of two different sets of best-fitting parameters for two individual LCs]{2012MNRAS.426.1291S}. Since the two high precision transit LCs available in literature were found to be inconsistent with each other our findings are from great importance for the determination of the system parameters.\\ Including the two publicly available data points, altogether 14 mid-transit times (the partial transit at epoch 517 was excluded) were used in the transit timing analysis. To investigate the error budget we ran a test to check the robustness of the transit times. We found, that if the ingress or egress, and hence the shape of the transit, is missing or show outlier and/or large scatter even the uncertainties determined with a wavelet-based MCMC may be underestimated. \\ We found no significant periodic signal in the O--C diagram. The strongest period at $\sim$26\,epochs has a FAP of 44.7\%. Hence, there is no evidence for TTV in the system. From our three-body-simulations, we can exclude even earth-mass perturbers in some resonance-orbital configurations. \\ Measurements of the Rossiter--McLaughlin effect found a misalignment between the stellar spin axis and the orbital axis of the planet which cannot be explained by planet-disc migration theories \citep[e.g.][]{2007ApJ...655..550G}. Another possible mechanism that may be responsible for such massive close-in planets is gravitational scattering by larger planets \citep{1996Natur.384..619W}. This planet--planet scattering scenario is one way to explain the properties of WASP-14\,b. \\ The significant eccentricity indicate either that the tides did not had sufficient time to influence the planetary orbit, which would support the planet--planet scattering theory or that the tidal effects in planetary systems are weaker than expected. Long-term follow-up studies of WASP-14 will add stricter constraints on these theories.


\section*{Acknowledgements}

We would like to thank H. Gilbert, S. Fiedler, I. H\"{a}usler, A. Reithe, and W. Rammo for participating in some of the observations at the University Observatory Jena.\\ SR is currently a Research Fellow at ESA/ESTEC. SR would like to thank DFG for support in the Priority Programme SPP 1385 on the `First Ten Million Years of the Solar system' in projects NE 515/33-1 and -2. MM acknowledge DFG for support in programme MU2695/13-1. AB would like to thank DFG for support in project NE 515/32-1. TE and LT would like to thank the DFG for support from the SFB-TR 7. CM acknowledges support from the DFG through grant SCHR665/7-1. MV would like to thank the projects APVV-0158-11 and VEGA 2/0143/14. GM and MV would like to thank the European Union in the Framework Programme FP6 Marie Curie Transfer of Knowledge project MTKD-CT-2006-042514 for support. TG acknowledges support from Bilim Akademisi -- The Science Academy, Turkey under the BAGEP programme. TG has been supported in part by Istanbul University: Project number 39742. T100 observations were performed under the project 12CT100-388. We would like to acknowledge financial support from the Thuringian government (B 515-07010) for the STK CCD camera used in this project.

\bibliographystyle{mn2e}
\bibliography{literatur}


\label{lastpage}

\end{document}